\documentclass[twocolumn,showpacs,aps,prd,nobibnotes,nofootinbib,floatfix]{revtex4-2}
\usepackage{amsmath,amssymb,graphics,epsfig,subfigure}
\usepackage{color}
\usepackage{multirow}
\usepackage{booktabs}
\usepackage{changes}
\usepackage{footnote}

\usepackage{hyperref}
\hypersetup{colorlinks=true,linkcolor=blue,citecolor=magenta}
\usepackage{booktabs}
\usepackage{orcidlink}

\DeclareMathOperator{\arctanh}{arctanh}
\DeclareMathOperator{\Tr}{Tr}

\begin{document}

\title{Gravitational partition function under volume constraints}


\author{Shan-Ping Wu\orcidlink{0000-0001-6633-1054}$^a$$^b$}
    \author{Peng Cheng\orcidlink{0000-0001-5704-7668}$^c$}    
    \author{Shao-Wen Wei\orcidlink{0000-0003-0731-0610}$^a$$^b$}
    \email{weishw@lzu.edu.cn, corresponding author}

\affiliation{$^{a}$Lanzhou Center for Theoretical Physics, Key Laboratory of Theoretical Physics of Gansu Province, and Key Laboratory of Quantum Theory and Applications of MoE, Lanzhou University, Lanzhou, Gansu 730000, China\\
$^{b}$Institute of Theoretical Physics $\&$ Research Center of Gravitation, Lanzhou University, Lanzhou 730000, China\\
$^{c}$Center for Joint Quantum Studies and Department of Physics, School of Science, Tianjin University, Tianjin 300350, China}
\date{\today}

\begin{abstract}
    The Euclidean action provides a bridge between gravitational thermodynamics and the partition function. In this work, we further investigate the gravitational partition function under a fixed-volume constraint, generalizing the fixed-volume on-shell geometry in the massless case. Moving beyond this massless configuration, we construct solutions with nonvanishing mass functions, which give rise to a new class of volume-constrained Euclidean geometries (VCEGs). These geometries possess both a boundary and a horizon. However, closer inspection indicates that the boundary is not intrinsic, but rather artificially introduced and can be extended, leading to the extended volume-constrained Euclidean geometries (ECVEGs). The ECVEGs contain two horizons, each generically associated with a conical singularity. Their Euclidean action is given by one quarter of the sum of the areas of the two horizons. In general, the conical singularities at the two horizons cannot be simultaneously eliminated, except at a critical mass $m = m^*$, which defines the critical ECVEG. Configurations with unavoidable conical singularities are naturally interpreted as constrained gravitational instantons. An analysis of their contributions to the partition function, together with their topological properties, reveals a close analogy between the ECVEGs and the Euclidean Schwarzschild--de Sitter static patch. This suggests that the volume constraint effectively plays a role analogous to that of a cosmological constant in semiclassical quantum gravity.
    \end{abstract}


	\maketitle
	\flushbottom
	\section{Introduction}

    Black holes occupy a central place in modern theoretical physics, serving as a natural arena where gravity and quantum theory intersect. Seminal work has revealed that black holes emit radiation, now known as Hawking radiation~\cite{Hawking1975Particle}, and exhibit a range of thermodynamic properties~\cite{Bekenstein:1973BHE,Hawking:1976BHT,Hartle:1976Path,Gibbons1976Action}. In particular, they obey analogs of the four laws of thermodynamics~\cite{Bekenstein:1972SecondLaw,Bardeen:1973FourLaws,Bekenstein:1974GeneralizedSecondLaw} and can undergo phase transitions~\cite{Hawking:1982HP,Chamblin:1999tk,Chamblin:1999hg,Kubiznak:2012PV}, placing them on a similar footing with ordinary thermodynamic systems. Yet black hole thermodynamics also presents profound departures from conventional thermodynamics. Most notably, the entropy of a black hole is not directly related to its volume but instead scales with the area of its event horizon. This area law strongly suggests that the microscopic degrees of freedom of gravity are encoded on lower-dimensional surfaces, pointing toward the holographic nature of gravitational systems~\cite{tHooft:1993dmi,Susskind:1994vu,Maldacena:1997re,Witten:1998qj}.

    Along the way to black hole thermodynamics, the Euclidean path integral has proven to be a powerful and successful approach~\cite{Hartle:1976Path,Gibbons1976Action}. Within the framework of thermal quantum field theory, the gravitational partition function can be formulated as a path integral,
    \begin{equation}
    	\mathcal{Z}_{\beta}=\int \left[ \mathcal{D} g \right] \exp \left( -I_E[g] \right),
    \end{equation}
    where $I_E[g]$ denotes the Euclidean gravitational action evaluated on a metric $g$, and $\beta$ represents the inverse temperature associated with the Euclidean time periodicity. The path integral is taken over all metrics that are periodic in $\beta$ and satisfy the appropriate asymptotic fall-off conditions. In practice, evaluating this path integral is highly nontrivial. When fluctuations are small, however, the saddle-point approximation provides an effective approach, yielding
    \begin{equation}
    	\mathcal{Z} \simeq \exp \left( -I_E\left[ g_{\text{on-shell}} \right] \right),
    \end{equation}
    where $g_\text{on-shell}$ denotes a metric that satisfies the classical equations of motion, i.e., an on-shell geometry. From this partition function, one can systematically derive thermodynamic quantities such as energy, entropy, and free energy. When the Einstein--Hilbert action with a positive cosmological constant is considered, the Euclidean section obtained by Wick rotating the Lorentzian de Sitter (dS) static patch provides an on-shell saddle. Its contribution to the partition function is given by
    \begin{equation}
    	\mathcal{Z} \simeq \exp \left( \mathcal{A}_\mathcal{H}/ (4 \hbar G) \right),
    \end{equation}
    where $\mathcal{A}_\mathcal{H}$ denotes the area of the dS horizon. For geometries satisfying the field equations (or, more precisely, the Hamiltonian and momentum constraints), the corresponding action is typically determined by contributions from horizons or boundaries. This feature has been discussed extensively in the literature~\cite{Hayward:1990zm,Whiting:1988qr,Banihashemi:2022jys,Wu:2025jes} and will also be reviewed in Sec.~\ref{Sec_EuclideanFormalismIyerWaldEntropy}. In the Euclidean dS static patch, where the geometry is smooth and possesses no boundary other than the regularized horizon, the Euclidean action is proportional to the horizon area. While this structure is manifest within the gravitational path integral, its physical interpretation remains less transparent. Viewed from the Euclidean static patch, the geometry is compact and boundaryless. It can be formally regarded as having an artificial internal spherical boundary of radius $R_B$, with the limit $R_B \to 0$~\cite{Banihashemi:2022jys,Jacobson:2022FVPartitionFunction}. In this limit, the corresponding Brown--York Hamiltonian $H_\text{BY}$ vanishes. The partition function can then be expressed as a trace over the Hilbert space states, $\mathcal{Z} = \Tr \exp(-\beta H_\text{BY}) = \Tr I$. Hence, the dimension of the Hilbert space is directly given by the partition function, and is therefore proportional to $\exp(\mathcal{A}_\mathcal{H} / 4 \hbar G)$. This idea was advocated by Fischler~\cite{Fischler:2000} and Banks~\cite{Banks:2000fe}, who argued that, since the sphere partition function is an integral unconstrained by any boundary conditions, the quantity $\mathcal{Z}_\beta$ must represent the dimension of the Hilbert space of all states in the theory describing a spatial volume. Furthermore, from a holographic perspective, developments in dS/CFT suggest that this entropy reflects the number of quantum gravitational degrees of freedom associated with the cosmological horizon~\cite{Strominger:2001pn,Banks:2003ta,Dong:2018cuv,Arias:2019pzy,Arenas-Henriquez:2022pyh,Chang:2024voo,Chang:2025ays}.

    More recently, these ideas have been extended to the study of Einstein gravity under a fixed-volume constraint, with the corresponding partition function investigated in Ref.~\cite{Jacobson:2022FVPartitionFunction}. In this framework, the on-shell geometry is compact, featuring a horizon but no boundary, similar to pure dS spacetime. The contribution of such an on-shell geometry to the Euclidean action can be interpreted as the dimension of the Hilbert space of a spatial region with the topology of a ball and fixed proper volume, given precisely by one quarter of the horizon area. Extensions to higher-derivative theories have also been explored, including massive gravity~\cite{Tavlayan:2023FVHigherCurvatureTheory} and Lovelock gravity~\cite{Lu:2024FVLovelockTheory}, where the strict area law is replaced by the Wald entropy~\cite{Wald1993BHentropy}.

    In the fixed-volume framework, the on-shell solutions presented in Ref.~\cite{Jacobson:2022FVPartitionFunction} were obtained by solving the Tolman--Oppenheimer--Volkoff (TOV) equation~\cite{Tolman:1939jz,Vessot:1980zz,PoncedeLeon:2000pj,Pleyer:2024oyb} with a vanishing mass function, without considering nonzero mass functions or matter sources. In this work, we go beyond the massless case and construct a broader class of volume-constrained Euclidean geometries (VCEGs) with nonzero mass functions. These configurations are physically relevant for two reasons. First, geometries with a nonzero constant mass remain valid solutions to the volume-constrained gravitational field equations and contribute meaningfully to the path integral---for instance, the Schwarzschild black hole with nonzero mass function yields a nontrivial on-shell action. Second, to explore possible excitations under the volume constraint, we introduce a simple model that describes how the mass function and matter sources affect the resulting on-shell geometry. Notably, the vanishing-mass case closely resembles the Euclidean dS static patch. As we will show below, when the mass function becomes nonzero, the geometry naturally develops features analogous to the static patch of a Euclidean Schwarzschild--de Sitter (SdS) black hole.

    On the other hand, these volume-constrained solutions can also be interpreted as constrained gravitational instantons~\cite{Affleck:1980mp,Frishman:1978xs,Cotler:2020lxj,Cotler:2021cqa}. Although they are not vacuum solutions of the Einstein equations without the volume constraint, they contribute nontrivially to nonperturbative processes in quantum gravity. The volume parameter effectively labels different constraint sectors, allowing the total gravitational partition function to be viewed as a sum over instantons with varying volumes. In this work, we focus on Einstein gravity with a fixed-volume constraint and the associated VCEGs. We find that, for a constant nonzero mass function, the VCEGs admit a natural extension, which we call extended volume-constrained Euclidean geometries (EVCEGs). These extended geometries are off-shell and should be interpreted as mass-constrained gravitational instantons. In the de Sitter (dS) case, the only genuinely smooth instantons are the Euclidean dS static patch and the Nariai geometry, whereas generic Schwarzschild--de Sitter (SdS) configurations are more naturally viewed as constrained instantons with fixed mass~\cite{Draper:2022xzl,Morvan:2022ybp}. We also observe a close analogy between the EVCEGs with nonzero mass and the SdS black hole.

    We therefore regard the EVCEGs with nonzero mass as playing a role analogous to that of the SdS black hole in the study of gravitational instantons~\cite{Draper:2022xzl,Morvan:2022ybp}. As constrained instantons, their contributions to the Euclidean action are physically significant. It is thus natural to investigate how these geometries contribute to the action and the corresponding partition function. To analyze the contribution of geometries to the action, we employ the covariant phase space formalism~\cite{Wald1993BHentropy,Iyer1994SomeProperties,Iyer:1995Comparison,Seraj:2016cym,Compere:2019qed,Harlow:2019yfa,Guo:2024oey,Guo:2025ohn,Wu:2025jes} within the fixed-volume framework. This approach shows that, for configurations satisfying the Hamiltonian and momentum constraints, the dominant contributions to the action come from the horizon and the boundary. The horizon term reproduces the Wald entropy, while the boundary term typically corresponds to the energy. Importantly, this analysis is general and applies to a broad class of covariant metric theories of gravity beyond Einstein gravity. Once the Euclidean action is obtained, we examine its implications for the gravitational path integral, including the interpretation of ensembles and decay rates.

    The main purpose of this work is to extend the class of metric solutions in Einstein gravity under a fixed-volume constraint and to investigate their geometric properties, contributions to the gravitational partition function, and associated physical implications. The paper is organized as follows. In Sec.~\ref{Sec_MassExcitation}, we study the TOV equation with the effective matter source induced by the volume constraint and obtain the VCEGs. In Sec.~\ref{Sec_Extension}, we analytically extend the VCEGs via a suitable coordinate transformation and construct the EVCEGs. In Sec.~\ref{Sec_GeometricContribution}, using the covariant phase space formalism, we analyze the contributions of the EVCEGs to the Euclidean action and demonstrate how the horizon encodes the entropy. In Sec.~\ref{Sec_ConstrainedInstantons}, we review the concept of constrained gravitational instantons, interpret the EVCEGs as mass-constrained instantons, and discuss their role in the gravitational partition function as well as the decay rates of the massless EVCEGs. Finally, Sec.~\ref{Sec_Discussion} summarizes our main findings and highlights the close connection between the EVCEGs and the SdS black hole.

	\noindent  \textbf{Notation}:
    Throughout this paper, we set $G=c=\hbar=1$, unless otherwise stated.
	The numerically invariant Levi-Civita symbol is denoted by $\varepsilon _{a_1...a_d}$, taking values $0$, $1$, or $-1$. We also introduce
	\begin{align*}
		\epsilon _{a_1...a_d}\equiv&\sqrt{\left| g \right|}\varepsilon _{a_1...a_d},
		\\
		\left( \mathrm{d}^nx \right) _{a_1...a_{\left( d-n \right)}}\equiv &\frac{1}{\left( d-n \right) !n!}\varepsilon _{a_1...a_{\left( d-n \right)}a_{\left( d-n+1 \right)}...a_d}
        \\
        &\mathrm{d}x^{a_{\left( d-n+1 \right)}}\land ...\land \mathrm{d}x^{a_d}.
	\end{align*}
    For simplicity we omit explicit references to the static patch. Unless stated otherwise, Euclidean dS spacetime and Euclidean SdS black hole are understood to refer exclusively to their respective static patches.



	\section{Metric solutions in finite volume}\label{Sec_MassExcitation}
	
	When a fixed spatial volume is imposed, the $D$-dimensional (or equivalently, $d+2$-dimensional) Euclidean action of Einstein gravity can be written as
	\begin{equation}
    \begin{split}
        I_E=&-\frac{1}{16\pi}\int_M{\sqrt{g}R\mathrm{d}^Dx}
        \\
        &-\int{\mathrm{d}\tau \lambda (\tau )}\left( \int_{\Sigma _{\tau}}{\sqrt{\gamma}\mathrm{d}^{D-1}x}-V \right) 
        \\
        &-\frac{1}{8\pi}\int_{\partial M}{\sqrt{h}K\mathrm{d}^{D-1}x},
    \end{split}		
		\label{eq_EuclideanAction}
	\end{equation}
	where the second term decomposes the integration over the manifold $M$ into an integral along the Euclidean time direction $\tau$ and an integral over the spacelike hypersurfaces $\Sigma_\tau$ at fixed $\tau$. Here, $\gamma_{ab}$ is the induced metric on $\Sigma_\tau$, and the Lagrange multiplier $\lambda(\tau)$ enforces the volume constraint by fixing the spatial volume $V$ on each slice. The third term is the Gibbons--Hawking--York (GHY) boundary term, where $h_{ab}$ is the induced metric on $\partial M$ and $K_{ab}$ is the trace of the extrinsic curvature. Varying this action with respect to the metric $g^{ab}$ and the Lagrange multiplier $\lambda(\tau)$ yields the equations of motion,
	\begin{equation}
		R_{ab}-\frac{1}{2}g_{ab}R=\frac{8\pi \lambda}{N}\gamma _{ab},\quad V=\int_{\Sigma _{\tau}}{\sqrt{\gamma}\mathrm{d}^{D-1}x},
		\label{eq_EOMforVC}
	\end{equation}
	where $N$ is the lapse function, which can be written explicitly as $N = \sqrt{g/\gamma}$. In the spherically symmetric and static case, the field equations admit metric solutions~\cite{Jacobson:2022FVPartitionFunction},
	\begin{equation}
    \begin{split}
        ds^2&=\frac{1}{4R^2}\left( R^2-r^2 \right) ^2d\tau ^2+dr^2+r^2d\Omega _{d}^{2}, 
        \\
        V &= \int_0^{R} r^{d}dr d\Omega _{d}.
    \end{split}		
		\label{eq_ds2meq0}
	\end{equation}
	The geometry admits a timelike Killing vector $\xi = \partial_\tau$ with surface gravity $\kappa = \sqrt{\nabla_a \xi_b \nabla^a \xi^b / 2} = 1$, which corresponds to an inverse temperature of period $\beta = 2\pi$. Since the manifold is compact and contains a horizon (at $r = R$) but no boundary, its contribution to the partition function is $\mathcal{Z} = \exp\left(\mathcal{A}/4\right)$, where $\mathcal{A}$ denotes the horizon area~\footnote{For a detailed discussion, see Sec.~\ref{Sec_EuclideanFormalismIyerWaldEntropy}.}. However, the corresponding solution is characterized by a vanishing mass function $M(r)=0$ and thus does not represent the most general case. In what follows, we will construct more general solutions.
	
	As a starting point, we consider the gravitational field equations coupled to a perfect fluid, which take the form
    \begin{align}
        & R_{ab}-\frac{1}{2}g_{ab}R=8\pi T_{ab}, \label{eq_effEOM}
        \\
        & T_{ab}=-\rho \left( r \right) n_an_b+p\left( r \right) \gamma _{ab}, \label{eq_Tab}
    \end{align}
    where $\rho(r)$ denotes the energy density of the perfect fluid, and $p(r)$ is the corresponding pressure. We adopt the spherically symmetric ansatz
    \begin{equation}
        ds^2=N\left( r \right) ^2d\tau ^2+\left( 1-\frac{16\pi}{d\Omega _d}\frac{M\left( r \right)}{r^{d-1}} \right) ^{-1}dr^2+r^2d\Omega _{d}^{2},
        \label{eq_Ansatz}
    \end{equation}
    where $M(r)$ denotes the mass function and $N(r)$ is the lapse function, both of which are to be determined. For convenience, and without loss of generality, we assume $N(r)>0$. Because of spherical symmetry, the angular components of the field equations are equivalent. It therefore suffices to consider the $\tau \tau$-component, the $rr$-component, and one representative angular component of Eq.~\eqref{eq_effEOM}. Thus, only three independent equations need to be analyzed. From $\tau \tau$- and $rr$-components, one readily obtains
    \begin{align}
        M^{\prime}(r)=&\rho \left( r \right) \Omega _dr^d,
        \label{eq_EqforM}
        \\
        N^{\prime}\left( r \right) =&-\frac{8\pi N\left( r \right) \left( \left( d-1 \right) M\left( r \right) +r^{d+1}\Omega _dp\left( r \right) \right)}{16\pi rM\left( r \right) -d\Omega _dr^d}.
        \label{eq_EqforN}
    \end{align}
    It is clear that, once the density $\rho(r)$ and pressure $p(r)$ are specified, Eqs.~\eqref{eq_EqforM} and \eqref{eq_EqforN} uniquely determine the metric. In general, however, the resulting solution does not satisfy the angular components of the field equations. This implies that $\rho(r)$ and $p(r)$ are subject to an additional condition arising from the angular components of the field equations. When incorporated into Eqs.~\eqref{eq_EqforM} and \eqref{eq_EqforN}, this condition is exactly equivalent to the conservation of the energy-momentum tensor, $\nabla_a T^{ab} = 0$. Of these, only the $r$-component is nontrivial and reads,
    \begin{equation}
         \left( \rho \left( r \right) +p\left( r \right) \right) \frac{N^{\prime}\left( r \right)}{N\left( r \right)}+p^{\prime}\left( r \right) =0.
         \label{eq_EqConstraint}
    \end{equation}
    Thus, for the three equations given by Eqs.~\eqref{eq_EqforM}, \eqref{eq_EqforN}, and \eqref{eq_EqConstraint}, there are four unknown functions: $N(r)$, $M(r)$, $\rho(r)$, and $p(r)$. By specifying one of these functions appropriately, or by imposing an additional suitable constraint, both the gravitational metric and the corresponding energy-momentum tensor can be determined.

    When the volume constraint, Eq.~\eqref{eq_EOMforVC}, is imposed, a natural relation arises,
    \begin{equation}
    	p(r) = \frac{\lambda}{N(r)},
    \end{equation}
    which fully determines the solutions. To proceed, imposing this relation simplifies Eq.~\eqref{eq_EqConstraint}, which becomes
    \begin{equation}
        \frac{\rho \left( r \right) p^{\prime}\left( r \right)}{p\left( r \right)}=0         \quad \text{or equivalently} \quad
         \frac{M^\prime \left( r \right) N^{\prime}\left( r \right)}{N\left( r \right)}=0.
         \label{eq_EqConstraint2}
    \end{equation}
    From this, we see that there are two possible types of solutions: first, the case $p'(r) = 0$, or equivalently $N'(r) = 0$; and second, the case $\rho(r) = 0$, or equivalently $M'(r) = 0$. These two cases correspond to distinct classes of solutions.

    \subsection{Case I: $p'(r) = 0$}\label{Sec_ppr=0}
    In this case, we have $p'(r) = 0$, which is equivalent to $N'(r) = 0$. Consequently, Eqs.~\eqref{eq_EqforM}, \eqref{eq_EqforN}, and \eqref{eq_EqConstraint} reduce to algebraic equations. The lapse function becomes constant, specifically $N(r) = N_0$, where $N_0$ is a constant. The corresponding metric solution reads
    \begin{equation}
        ds^2=N_{0}^{2}d\tau ^2+\frac{1}{1-\frac{16\pi \rho _0}{d\left( d+1 \right)}r^2}dr^2+r^2d\Omega _{d}^{2},
        \label{eq_ds2SS}
    \end{equation}
    with the energy density and pressure given by
    \begin{equation}
        \rho \left( r \right) =\rho _0,\quad p\left( r \right) =\frac{\lambda}{N_0},
        \label{eq_rhopC}
    \end{equation}
    where
    \begin{equation}
        \rho _0=-\frac{d+1}{d-1}\frac{\lambda}{N_0}.
    \end{equation}
    When the constant energy density satisfies $\rho_0 > 0$, the metric is confined to the radial region $r \in [0,r_c)$, where
    \begin{equation}
        r_c=\sqrt{\frac{d\left( d+1 \right)}{16\pi \rho _0}}.
    \end{equation}
    At $r = r_c$, the metric appears to diverge; however, this is merely a coordinate singularity. Introducing a new radial coordinate $\eta$,
    \begin{equation}
        \eta =\arcsin{\left( \sqrt{\frac{16\pi \rho _0}{d\left( d+1 \right)}}\,\,r \right)},
    \end{equation}
    the metric becomes manifestly regular,
    \begin{equation}
        ds^2=N_{0}^{2}d\tau ^2+\frac{d\left( d+1 \right)}{16\pi \rho _0}\left( d\eta ^2+\sin ^2\eta d\Omega _{d}^{2} \right).
    \end{equation}
    Note that $r \in [0, r_c)$ corresponds to $\eta \in [0, \pi/2)$. In the $\eta$ coordinate, the geometry admits a natural extension, with $\eta \in [0, \pi]$. In this form, it is clear that the spatial section corresponds to a $(d+1)$-sphere with volume,
    \begin{equation}
        V=\Omega _{d+1}\left( \frac{d\left( d+1 \right)}{16\pi \rho _0} \right) ^{\left( d+1 \right) /2},
    \end{equation}
    and the full geometry is given by the product space $\mathbb{S}^1 \times \mathbb{S}^{d+1}$. If $\rho_0 < 0$, the spacetime is not confined to a finite radial region. By applying a coordinate transformation similar to the one discussed above, one finds that the spatial section becomes a hyperbolic space. Physically, even in the presence of the volume constraint, a negative energy density allows the spacetime to extend indefinitely.

    Let us briefly examine the energy conditions. For the energy density and pressure specified in Eq.~\eqref{eq_rhopC}, one finds that if $\lambda<0$, then
    \begin{equation}
    \begin{split}
        &\rho =\rho _0>0,\quad \rho +p=\frac{2}{d+1}\rho _0,
        \\
        &\rho +\left( d+1 \right) p=-d\rho _0<0,\quad \rho >\left| p \right|.
    \end{split}
    \end{equation}
    It then follows that the weak and dominant energy conditions are satisfied, while the strong energy condition is violated. However, in the present setup, the volume constraint has been effectively interpreted as a matter pressure. In fact, the additional excitation under consideration carries only energy and no pressure. Therefore, when $\lambda<0$, all three standard energy conditions are actually satisfied.

    \subsection{Case II: $\rho(r) = 0$}\label{Sec_rhoequal0}
    In this case, we have $\rho(r) = 0$, which is equivalent to $M'(r) = 0$. Since the energy density vanishes, we have
    \begin{equation}
        \rho \left( r \right) =0, \quad p\left( r \right) =\frac{\lambda}{N\left( r \right)}.
    \end{equation}
    Equation~\eqref{eq_EqforM} then implies that the function $M$ takes a constant value. To avoid a naked singularity at $r=0$, we assume this constant to be positive. For convenience, we parametrize it as
    \begin{equation}
        M\left( r \right) =\frac{d\Omega _d}{16\pi}m^{d-1},
    \end{equation}
    where $m$ is a constant with the dimension of length, assumed to be non-negative. For clarity and consistency in the subsequent discussion, we refer to $m$ as the mass parameter, and to $M$ as the mass constant (or mass function). Under this parametrization, the metric ansatz \eqref{eq_Ansatz} takes the form
    \begin{equation}
        ds^2=N\left( r \right) ^2d\tau ^2+\left( 1-\frac{m^{d-1}}{r^{d-1}}\right)^{-1} dr^2+r^2d\Omega _{d}^{2}.
        \label{eq_ds2Nr}
    \end{equation}
Since we are considering a Euclidean geometry here, the metric must be positive definite. Therefore, we impose this condition, which requires $g_{rr}>0$ and consequently restricts the geometry to the region $r \ge m$. The remaining function $N(r)$ is then determined by Eq.~\eqref{eq_EqforN}, which becomes
    \begin{equation}
        N^{\prime}\left( r \right) =\frac{d\left( d-1 \right) m^{d-1}N\left( r \right) +16\pi \lambda r^{d+1}}{2dr\left( r^{d-1}-m^{d-1} \right)}.
        \label{eq_Npr}
    \end{equation}
    Since we are considering the region $r \ge m$ and $N(r) \ge 0$, one immediately observes that if $\lambda > 0$, the function $N(r)$ is monotonically increasing. Consequently, $N(r)$ admits no zeros, and the geometry remains open. On the other hand, for $\lambda < 0$, Eq.~\eqref{eq_Npr} implies that the asymptotic behavior of the solution at large $r$ is given by
    \begin{equation}
        N\left( r \right) \sim \lambda r^2 <0.
    \end{equation}
    This implies that $N(r)$ must necessarily vanish at some finite radius greater than $m$, denoted by $r = r_c$. Substituting $r=r_c$ into Eq.~\eqref{eq_Npr}, one finds $N'(r_c) < 0$. Therefore, for a continuous solution, $N(r)$ possesses a unique zero, and the physically admissible domain is restricted to the interval $[m, r_c]$. From Eq.~\eqref{eq_Npr}, together with the condition $N(r_c) = 0$, one can solve for $N(r)$ as
    \begin{equation}
    \begin{split}
         N\left( r \right) &=\frac{-8\pi \lambda m^2}{d}\sqrt{1-\frac{m^{d-1}}{r^{d-1}}}C\left( r \right),
         \\
        C\left( r \right) &=\int_{m/r_c}^{m/r}{\frac{1}{x^3\left( 1-x^{d-1} \right) ^{3/2}}dx}.
    \end{split}
    \label{eq_SolforNC}
    \end{equation}
    Note that $m \le r \le r_c$, so that the integration range in the second expression is constrained by $m/r_c \le m/r \le 1$. It is further worth noting that as $r\rightarrow m$, the integral behaves asymptotically as $C(r) \sim (1-m/r)^{-1/2}$. Although this suggests a divergence, the prefactor in the expression for $N(r)$ precisely cancels this divergence, ensuring that $N(r)$ remains finite at $r = m$. Another singularity of interest arises in $g_{rr}$ at $r = m$, which, as in the case discussed in Sec.~\ref{Sec_ppr=0}, is merely a coordinate singularity and can be removed via an appropriate coordinate transformation.

    One observes that the resulting geometry shares certain similarities with the global black hole geometry; however, there are crucial differences. In particular, although $g^{rr}$ vanishes at $r = m$, the metric function $g_{tt} = N(r)^2$ does not. Upon continuation to Lorentzian signature, the region $r < m$ would develop two timelike directions, one along the original time coordinate and the other along the radial direction. We therefore regard this sector as unphysical and do not extend the geometry into $r < m$. On the other side, the region $r > r_c$, where $N(r)$ becomes negative, cannot be analytically continued. Recall that the original definition $p(r) = \lambda / |N(r)|$ can be simplified to $p(r) = \lambda / N(r)$ under the assumption $N(r) > 0$. Once $N(r) < 0$, this assumption breaks down, Eq.~\eqref{eq_Npr} undergoes a substantial change, and the continuation of $N(r)$ no longer satisfies the field equations. Consequently, the spacetime under consideration is strictly restricted to the domain $m \le r \le r_c$.

    It is important to note that at $r = r_c$ the geometry exhibits two distinct types of singular behavior. The first is a conical singularity. Since the Euclidean time $\tau$ is periodic, a conical defect generally arises at $r = r_c$. However, this singularity can be removed by an appropriate choice of the period of $\tau$. Secondly, there is also a curvature singularity at this point, which cannot be removed. Taking the trace of Eq.~\eqref{eq_EOMforVC}, one finds
    \begin{equation}
        R = -\frac{16\pi \lambda (d+1)}{d}\frac{1}{N}.
    \end{equation}
    This implies that the Ricci scalar diverges at the horizon where $N \to 0$. Nevertheless, this divergence is sufficiently mild in the sense that it does not lead to a divergence of the Euclidean action. More precisely, although the curvature scalar behaves as $R \sim 1/N$, the integration measure provides a compensating factor through $\sqrt{g}$, so that the action remains finite. Therefore, this curvature singularity is integrable and does not invalidate the saddle-point geometry. Such behavior is not unexpected in constrained Euclidean gravitational systems, and similar features have also been discussed in Ref.~\cite{Jacobson:2022FVPartitionFunction}.
    
    By varying the Lagrange multiplier $\lambda$, one imposes the volume constraint. The corresponding constrained volume is obtained by integrating the spatial part of the metric over the radial interval from $r = m$ to $r = r_c$, yielding,
    \begin{equation}
        V=\Omega _dm^{d+1}\int_{m/r_c}^1{\frac{1}{\left( 1-x^{d-1} \right) ^{1/2}x^{d+2}}dx}.
        \label{eq_V}
    \end{equation}
    It is straightforward to observe that, although the integrand exhibits a singularity at $x = 1$ (i.e., $r = m$), the divergence behaves as $(1-x)^{-1/2}$ and is therefore finite in the sense of the integral. Finally, as the last discussion in this section, let us examine the limit $m \rightarrow 0$. In this limit, the solution \eqref{eq_SolforNC} reduces to that of Ref.~\cite{Jacobson:2022FVPartitionFunction}, after a suitable redefinition of the Euclidean time $\tau$.
    
    To summarize, we have examined spherically symmetric geometries in dimensions $d \geq 4$, characterized by $p(r)=\lambda/N(r)$. For these configurations to satisfy the field equations, the constraint \eqref{eq_EqConstraint2} must be imposed. Two distinct cases arise, namely $p'(r)=0$ and $\rho(r)=0$. In both cases, compact geometries exist only when $\lambda<0$; otherwise, the spatial volume diverges. In the first case, the energy density is nonvanishing and uniformly distributed throughout spacetime, giving rise to the geometry $\mathbb{S}^1 \times \mathbb{S}^{d+1}$. However, this configuration does not represent a genuine gravitational solution of the action \eqref{eq_EuclideanAction}. By contrast, in the second case the energy density vanishes, while the mass function $M(r)$ may remain nonzero. The resulting geometry constitutes a genuine gravitational solution of Eq.~\eqref{eq_EuclideanAction}, thereby extending the class of configurations considered in Ref.~\cite{Jacobson:2022FVPartitionFunction}.

    In the following, we focus on the solution discussed in Sec.~\ref{Sec_rhoequal0}. For convenience, we refer to this solution of Einstein gravity subject to a fixed-volume constraint as VCEG for short. The configuration with a vanishing mass constant ($M = 0$) will be referred to as the massless VCEG, whereas those with a nonzero mass constant ($M \neq 0$) will be termed massive VCEG.

   \section{Extension of the volume constrained Euclidean geometry}\label{Sec_Extension}

   In the preceding section, we analyzed a class of Euclidean geometries, the massive VCEGs, which arise when the mass parameter $m$ (or equivalently the mass constant $M$) is nonzero. Compared with the massless VCEGs, a nonzero $m$ introduces a cutoff of the spacetime at a finite radius. As discussed in Sec.~\ref{Sec_rhoequal0}, the region $r<m$ is unphysical, and one may therefore introduce a cutoff surface at $r=m$ to serve as the boundary. At this surface, the boundary geometry takes the form of a direct product $\mathbb{S}^1 \times \mathbb{S}^d$, with respective radii $N(m)$ and $m$, forming a finite throat. This suggests that the spacetime does not genuinely terminate at $r=m$; rather, the boundary at $r=m$ should be regarded as an artificial construct. Furthermore, as shown in Eq.~\eqref{eq_SolforNC}, the derivative of $N(r)$ diverges at $r=m$, indicating that $r$ is not a suitable coordinate in this region. These observations motivate the introduction of an alternative coordinate system in which the metric \eqref{eq_ds2Nr} is better behaved and admits a smooth extension.

   To achieve this, we introduce the coordinate,
    \begin{equation}
    \begin{split}
        &\cos ^2\chi =\frac{m^{d-1}}{r^{d-1}},\quad \text{or equivalently} 
        \\
        & \chi =\arccos \left( \left( m/r \right)^{\left( d-1 \right) /2} \right) .
    \end{split}        
        \label{eq_coordtran}
    \end{equation}
    The range $r \in [m, r_c]$ is then mapped to $\chi \in [0, \chi_c]$, where
    \begin{equation}
        \chi _c=\mathrm{arc}\cos \left( \left( m/r_c \right) ^{\left( d-1 \right) /2} \right) ,
    \end{equation}
    establishing a one-to-one correspondence between the two variables. In terms of coordinate $\chi$, the metric
    \begin{equation}
    \begin{split}
        ds^2=&N\left(  \chi \right) ^2 d\tau ^2+\frac{4m^2}{\left( d-1 \right) ^2}\cos \left( \chi \right) ^{-2\left( d+1 \right) /\left( d-1 \right)}d\chi ^2
        \\
        &+r\left( \chi \right) ^2 d\Omega _{d}^{2},
    \end{split}        
        \label{eq_newds2}
    \end{equation}
   with lapse function $N(\chi) \equiv N(r(\chi))$, despite the slight abuse of notation. Importantly, the coordinate transformation is technically degenerate at $r=m$, since $\partial_\chi r(\chi)|_{\chi=0}=0$. However, this very feature eliminates the divergence that was present in the radial component of the metric. Moreover, in the new coordinate $\chi$, the factor $\sqrt{1 - m^{d-1}/r^{d-1}}$ in the lapse function $N(r)$ is replaced by $\sin\chi$, thereby avoiding coordinate singularities. We demonstrate this explicitly in the cases of $D=4$ and higher dimensions.

   Since the range of $r$ is $[m, r_c]$, with $r = m$ representing a boundary, the corresponding range of $\chi$ is $[0, \chi_c]$, where $\chi = 0$ likewise serves as a boundary. In the original metric \eqref{eq_ds2Nr}, the restriction $r \geq m$ ensures the positive definiteness of the metric, as well as a well-defined causal structure upon Lorentzian continuation. By contrast, in the metric \eqref{eq_newds2}, the coordinate $\chi$ can be smoothly extended across $\chi = 0$, as a direct consequence of the transformation in Eq.~\eqref{eq_coordtran}. Nevertheless, this transformation also implies that $r \geq m$ holds for all $\chi$, so that no unphysical region is introduced. As will be discussed below, the original geometry defined on $\chi \in [0,\chi_c]$ can be extended to $\chi \in [\chi_e,\chi_c]$, where $\chi_e \in [-\pi/2,0]$ is determined by the condition $N(\chi_e) = 0$. The resulting geometry no longer contains a truncated boundary, but instead features two horizons at $\chi = \chi_e$ and $\chi = \chi_c$, closely resembling the structure of the SdS black hole. We refer to this extended configuration as the EVCEG for short. Furthermore, from the line element \eqref{eq_newds2}, $r$ admits a natural interpretation as the radius of the spherical section $\mathbb{S}^d$. In addition, the transformation \eqref{eq_coordtran} implies that $r(\chi)$ attains its minimum at $\chi = 0$, indicating that $\chi = 0$ (equivalently $r = m$) corresponds to a wormhole-like throat.

   \subsection{The four-dimensional case}

\begin{figure}[h]
        \centering
        \includegraphics[width=6cm]{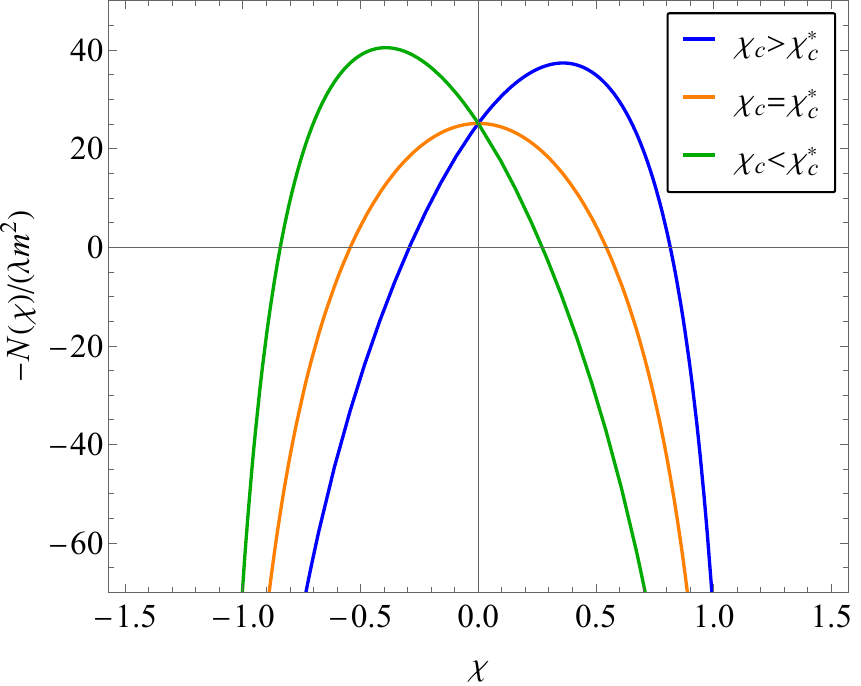}
        \caption{Profile of the lapse function $N(\chi)$. The curves with different colors correspond to different parameter choices: the blue curve represents the case $\chi_c>\chi_c^*$, the orange curve corresponds to $\chi_c = \chi_c^*$, and the green curve denotes $\chi_c < \chi_c^*$. The critical value $\chi_c^*$ is defined by the condition $H(\chi_c^*) = 0$.}
       \label{Fig_chiN}
    \end{figure}

\begin{figure}[h]
    \begin{center}
			\subfigure[ $\chi_c$ or $\chi_e$ vs. $\chi_c$
			\label{Fig_chicchie}]{\includegraphics[width=6 cm]{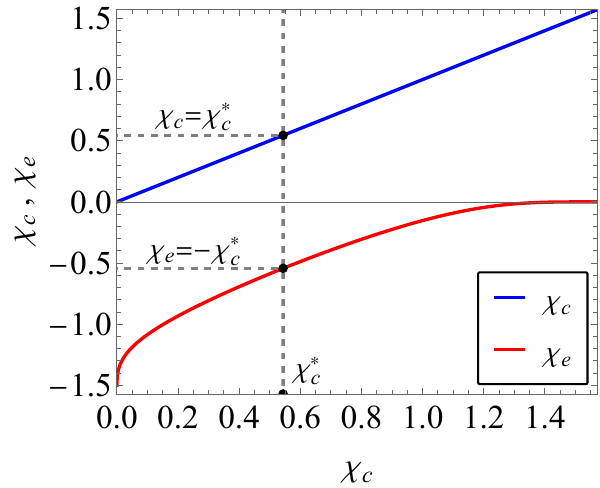}}
            \quad
			\subfigure[ $m$ vs. $\chi_c$
			\label{Fig_chicm}]{\includegraphics[width=6 cm]{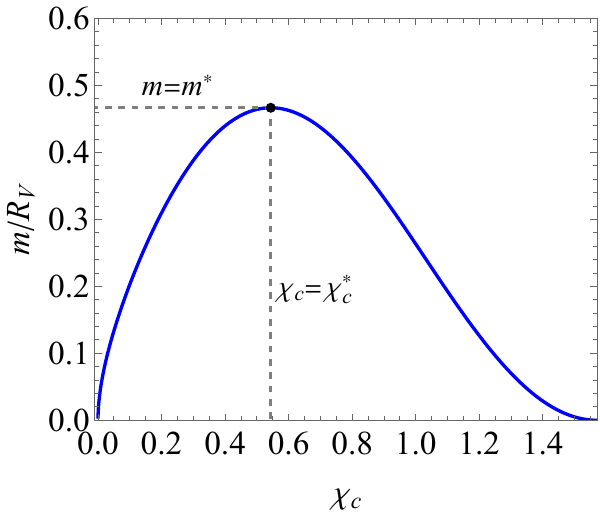}}
	\end{center}
        \caption{Relations among the characteristic parameters of the metric, $m$, $\chi_e$, and $\chi_c$. (a) The dependence of $\chi_c$ and $\chi_e$ on $\chi_c$. The blue and red curves correspond to $\chi_c$ and $\chi_e$, respectively, where $\chi_c$ and $\chi_e$ denote the two roots of the lapse function $N(\chi)$. (b) The relation between the mass parameter $m$ and $\chi_e$. The mass parameters reaches its maximal value $m^*$ when $\chi_c = \chi_c^*$.}
    \end{figure}

   We now turn to the four-dimensional case. Starting from the integral expression for the lapse function $N(r)$ in Eq.~\eqref{eq_SolforNC}, and performing the coordinate transformation $\cos^2\chi = m/r$, we obtain
   \begin{equation}
   \begin{split}
        N\left( \chi \right) &= -\pi \lambda m^2\sin \chi 
        \\
        &\left( -15\arctanh \left( \sin \chi \right) +\frac{15\cos^4\chi-5\cos^2\chi-2}{\sin \chi \cos^4\chi } \right.
        \\
        &\left.+15\arctanh \left( \sin  \chi _c \right) -\frac{15 \cos^4\chi_c -5\cos^2\chi_c-2}{\sin \chi _c \cos^4\chi_c } \right),
   \end{split}
   \label{eq_Nchi4}
   \end{equation}
   where $\chi_c = \arccos(\sqrt{m/r_c})$ denotes the largest root of $N(\chi)$. The behavior of $N(\chi)$ is illustrated in Fig.~\ref{Fig_chiN}. We observe that $N(\chi)$ generally possesses two zeros: the positive one at $\chi_c$ and another negative one at $\chi_e$. Moreover, $N(\chi)$ is not generically symmetric under $\chi \to -\chi$. This structure can be seen in Eq.~\eqref{eq_Nchi4}, where the first line of the bracketed expression is odd in $\chi$, while the second line is an even (constant) term. Only when the constant term vanishes does $N(\chi)$ become symmetric. In this case, the corresponding root is denoted as $\chi_c^*$, with the second root located at $\chi_e = -\chi_c^*$. Numerically, we obtain $\chi_c^* \approx 0.54$. Moreover, Eq.~\eqref{eq_Nchi4} implies that once $\chi_c$ is fixed, $\chi_e$ is determined uniquely; their relation is plotted in Fig.~\ref{Fig_chicchie}. A close inspection shows that the relation between $\chi_e$ and $\chi_c$ is nonlinear. In particular, as $\chi_c \to 0$ one finds $\chi_e \to -\pi/2$, so the extended coordinate range approaches $(-\pi/2,0)$. Conversely, as $\chi_c \to \pi/2$ one has $\chi_e \to 0$, and the range approaches $(0,\pi/2)$.

\begin{figure}[h]
        \centering
        \includegraphics[width=7cm]{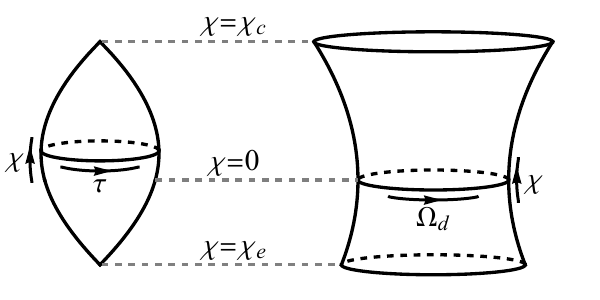}
        \caption{Schematic illustration of the EVCEG. The left panel shows the Euclidean section in the $(\tau, \chi)$ plane, forming a double-cone structure with tips at $\chi = \chi_e$ and $\chi = \chi_c$, which define the allowed range $\chi_e < \chi < \chi_c$. The right panel depicts the $(\chi, \Omega_{d})$ section, where the spherical radius reaches its minimum at $\chi = 0$, corresponding to the throat of the geometry with radius equal to the mass parameter $m$. The maximum of the lapse function $N(\chi)$ lies above the throat, as illustrated here for $\chi_c > \chi_c^*$. For $\chi_c = \chi_c^*$ or $\chi_c < \chi_c^*$, the extremum coincides with or lies below the throat, respectively.}
       \label{Fig_sketch}
    \end{figure}

    From a geometric perspective, the extended solution is compact and contains no boundary. In this case, the geometry admits a Euclidean time direction, and since $N(\chi_e) = N(\chi_c) = 0$ at both $\chi = \chi_e$ and $\chi = \chi_c$, the spacetime features two horizons. To avoid conical singularities, each horizon imposes a periodicity on the Euclidean time coordinate, yielding the corresponding inverse temperatures
    \begin{equation}
        \beta _c=\frac{-4\pi m}{N^{\prime}\left( \chi _c \right) \cos ^3\chi _c},\quad \beta _e=\frac{4\pi m}{N^{\prime}\left( \chi _e \right) \cos ^3\chi _e},
        \label{eq_betace}
    \end{equation}
    respectively. In general, these two periods are unequal, except when $\chi_c = \chi_c^*$, which indicates that the geometry typically contains at least one conical singularity. Along the radial direction, each value of $r$ defines a spherical section $\mathbb{S}^{2}$, with the radius attaining its minimum $r = m$ at $\chi = 0$, thereby forming a throat reminiscent of a wormhole. Altogether, this geometry resembles that of a Euclidean dS black hole, while simultaneously exhibiting wormhole-like features, bounded between the two horizons. A schematic illustration is shown in Fig.~\ref{Fig_sketch}.

    In addition to the equations of motion, there exists a further constraint, namely the volume constraint,
   \begin{equation}
   \begin{split}
       &V=\frac{\pi}{6}m^3
       \\
       &\left( 15 \arctanh \left( \sin \chi _c \right) +\frac{\sin \chi _c}{\cos ^6\chi _c}\left( 15\cos ^4\chi _c+10\cos ^2\chi _c+8 \right) \right.
       \\
      &\left. -15 \arctanh \left( \sin \chi _e \right) -\frac{\sin \chi _e}{\cos ^6\chi _e}\left( 15\cos ^4\chi _e+10\cos ^2\chi _e+8 \right) \right).
   \end{split}
   \label{eq_Vcon4}
   \end{equation}
   The volume constraint relates $m$, $\chi_e$, and $\chi_c$. Since $\chi_e$ is determined by $\chi_c$, it follows that $m$ is also fixed once $\chi_c$ is specified, as illustrated in Fig.~\ref{Fig_chicm}. Furthermore, from the periodicity of $\tau$ given in Eq.~\eqref{eq_betace}, one sees that $\lambda$, appearing as the overall coefficient of $N(r)$, can be used to adjust the relation between the conical deficit and the Euclidean time period, and can effectively be absorbed into the coordinate $\tau$. Consequently, the EVCEG in this case is essentially determined by the single parameter $\chi_c$.

   On the other hand, viewed in reverse, when the mass parameter $m$ is specified, the values of $\chi_e$ and $\chi_c$ can be obtained by solving Eq.~\eqref{eq_Vcon4} together with the condition $N(\chi_e)=0$. A noteworthy feature of these solutions is the presence of an underlying $\mathbb{Z}_2$ symmetry: if $(\chi_c,\chi_e)=(\hat{\chi}_c,\hat{\chi}_e)$ is a solution, then $(\chi_c,\chi_e)=(-\hat{\chi}_e,-\hat{\chi}_c)$ is also a solution. Moreover, one finds
   \begin{equation}
       \left. N\left( \chi \right) \right|_{\chi _c=\hat{\chi}_c}=\left. N\left( -\chi \right) \right|_{\chi _c=-\hat{\chi}_e},
   \end{equation}
   and since the $rr$- and angular components of the metric in the present coordinate system are even functions of $\chi$, we conclude that the two cases $\chi_c=\hat{\chi}_c$ and $\chi_c=-\hat{\chi}_e$ are in fact degenerate. They correspond to the same geometry, related by the coordinate transformation $\chi \to -\chi$. As a consequence, the effective range of $\chi_c$ is restricted to $[0,\chi_c^*]$ or $[\chi_c^*,\pi/2]$, where $\chi_c^*$ is indicated in Fig.~\ref{Fig_chicchie}. Alternatively, the geometry can be parametrized by $m$, whose range is $[0,m^*]$, with $m^*$ marked in Fig.~\ref{Fig_chicm}. Nevertheless, for practical convenience, we shall often use $\chi_c$ to characterize the geometry, despite the possible degeneracy.
   \subsection{Higher-dimensional case}

   For dimensions higher than four, a similar analysis can be carried out. However, under the volume constraint, the solutions, VCEG, become significantly more complicated, as illustrated in Eq.~\eqref{eq_SolforNC}. Nevertheless, many qualitative properties of the solutions can still be inferred directly from the equations of motion. In terms of the coordinate transformation \eqref{eq_coordtran}, Eq.~\eqref{eq_Npr} (or equivalently the field equations) yields the differential equation for $N(\chi)$,
   \begin{equation}
   \begin{split}
       &N^{\prime}\left( \chi \right)=
       \\
       &\frac{d\left( d-1 \right) \cos \chi N\left( \chi \right) +16\pi \lambda m^2\cos \left( \chi \right) ^{-\left( d+3 \right) /\left( d-1 \right)}}{d\left( d-1 \right) \sin \chi}.
   \end{split}       
       \label{eq_Npchi}
   \end{equation}
   As in the analysis of $N(r)$, finiteness of the spatial volume requires $\lambda<0$. Near the endpoints $\chi\to \pm \pi/2$, one finds the asymptotic behavior
   \begin{equation}
   N\left( \chi \right) \sim \lambda \left( \cos \chi \right) ^{-4/\left( d-1 \right)}.
   \end{equation}
   Since the valid region of the geometry is defined by the condition $N(\chi) > 0$, the function $N(\chi)$ must necessarily possess at least two zeros. At such zeros, say at $\chi_0$ where $N(\chi_0)=0$, Eq.~\eqref{eq_Npchi} reduces to
   \begin{equation}
   N^{\prime}\left( \chi_0 \right) =\frac{16\pi \lambda m^2}{d\left( d-1 \right) \sin \chi_0}\cos \left( \chi_0 \right) ^{-\left( d+3 \right) /\left( d-1 \right)}.
   \end{equation}
   This expression shows that the sign of $N'(\chi_0)$ at a zero is fixed: for a zero located at $\chi_0>0$, one has $N'(\chi_0)<0$, while for a zero located at $\chi_0<0$, one has $N'(\chi_0)>0$. Consequently, the metric function $N(\chi)$ generally possesses exactly one positive root and one negative root. In analogy with the four-dimensional case, we denote the positive root by $\chi_c$ and the negative one by $\chi_e$. For the differential equation \eqref{eq_Npchi} with the boundary condition $N(\chi_c)=0$, we can obtain the lapse function as,
   \begin{equation}
   \begin{split}
       &N\left( \chi \right) =-\frac{16\pi \lambda m^2}{d\left( d-1 \right)}\sin \chi \left( H\left( \chi \right) -H\left( \chi _c \right) \right),
       \\
       &H\left( \chi \right) =\frac{\cos \chi}{\sin \chi}\,_2F_1\left( -\frac{1}{2}, -\frac{d+3}{2(d-1)}; \frac{1}{2}; -\tan ^2\chi \right),
   \end{split}
   \label{eq_Nchid}
   \end{equation}
   where $_2F_1$ denotes the Gauss hypergeometric function. Meanwhile, the volume constraint yields
   \begin{equation}
   \begin{split}
       V=\frac{2 m^d\Omega _d}{d-1}&\left( {}_2F_1\left( \frac{1}{2},\frac{2d}{d-1};\frac{3}{2};\sin ^2\chi _c \right) \sin \chi _c \right.
       \\
       & \; \left. -{}_2F_1\left( \frac{1}{2},\frac{2d}{d-1};\frac{3}{2};\sin ^2\chi _e \right) \sin \chi _e \right) .
   \end{split}
       \label{eq_Vcd}
   \end{equation}
   For $d=2$, this result reduces consistently to Eq.~\eqref{eq_Nchi4}, recovering the four-dimensional case. One observes that $H(\chi)$ is an odd function, while $N(\chi)$ is not even unless $H(\chi_c)=0$. A further analysis of the metric solution reveals results analogous to those found in the four-dimensional case, leading to several parallel conclusions. First, there exists a potential $\mathbb{Z}_2$ symmetry between $\chi_c$ and $\chi_e$, indicating a redundancy in characterizing the geometry through either root. There is a critical value $\chi_c^{}$ satisfying $H(\chi_c^{*}) = 0$, at which the function $N(\chi)$ becomes even. The metrics corresponding to $\chi_c \in (0, \chi_c^{*})$ and $\chi_c \in (\chi_c^{*}, \pi/2)$ are related by the coordinate reflection $\chi \to -\chi$. Second, from a geometric viewpoint, in higher dimensions the Euclidean spacetime contains two horizons, with the sphere radius attaining its minimum value $m$ at $\chi = 0$, reminiscent of a wormhole throat. 

   It is worth noting that when the mass parameter $m$ vanishes identically (equivalently $\chi_c=0$ or $\chi_c=\pi/2$), the metric degenerates in the corresponding components and thus becomes ill-defined. However, an asymptotic analysis shows that as $\chi_c\to\pi/2$ one has $m\to 0$ and $\chi_e\to 0$. Using the coordinate transformation \eqref{eq_coordtran}, the resulting geometry approaches the massless VCEG \eqref{eq_ds2meq0} (after an appropriate choice of $\lambda$ the two metrics coincide). This indicates that the massless EVCEG coincides with the massless VCEG itself; in other words, when the mass parameter vanishes, no further extension is required. On the other hand, when $m$ reaches its maximal value $m^*$, one can eliminate the conical singularity at the horizon by an appropriate redefinition of $\tau$. In this case, the massive EVCEG is referred to as the critical EVCEG. On the other hand, when $m$ reaches its maximal value $m^*$, the conical singularity at the horizon can be eliminated by an appropriate redefinition of $\tau$. In this case, the massive EVCEG is referred to as the critical EVCEG. For $0 < m < m^*$, we refer to the solutions as non-critical massive EVCEGs. For convenience, we summarize the corresponding nomenclature in Table~\ref{Table_EVCEGclassification}.

   \begin{table}[htbp]
   \caption{Classification of EVCEG configurations according to the mass parameter $m$.}
   \label{Table_EVCEGclassification}
   \begin{ruledtabular}
        \begin{tabular}{cc}
        Mass parameter $m$ & Name \\
        \hline
        $m = 0$ & Massless EVCEG \\
        \hline
        $0 < m < m^*$ & Non-critical massive EVCEG \\
        \hline
        $m = m^*$ & Critical EVCEG \\
        \end{tabular}
    \end{ruledtabular}
    \end{table}

\section{Geometric contributions to the Euclidean action}\label{Sec_GeometricContribution}

      \subsection{Euclidean formalism and Iyer-Wald entropy}\label{Sec_EuclideanFormalismIyerWaldEntropy}

    In the previous section, we discussed the gravitational solution associated with a nonzero constant mass function. As shown in Eq.~\eqref{eq_SolforNC} or \eqref{eq_Nchid}, the expression for $N(r)$ is rather complicated, which may lead to difficulties in evaluating the Euclidean action. To better and more transparently identify the contribution of the geometry to the Euclidean action \eqref{eq_EuclideanAction}, we now turn to a more general discussion based on the covariant phase space formalism.

    In the context of gravitational theories, general covariance constitutes the fundamental gauge symmetry. Within the covariant phase space formalism, such symmetries give rise to conserved surface charges associated with exact spacetime symmetries. This framework provides a direct derivation of the first law of black hole thermodynamics~\cite{Wald1993BHentropy,Iyer1994SomeProperties}, and furthermore naturally encodes the relation between Noether charges and the on-shell gravitational action~\cite{Wald1993BHentropy,Iyer:1995Comparison,Wu:2025jes}. When a volume constraint is imposed in the action, however, subtleties arise in the covariant phase space construction. Since the spatial volume of each spacelike slice is fixed, the original diffeomorphism invariance is partially broken. Nevertheless, the relation between Noether charges and the on-shell gravitational action can still be established. Related discussions can be found in the Supplemental Material of Ref.~\cite{Jacobson:2022FVPartitionFunction}. Here we present a parallel review, including explicit derivations and a more detailed analysis.

    To begin with, the bulk part of the Euclidean action \eqref{eq_EuclideanAction} for $D$-dimensional gravitational theories at finite volume can be expressed as
    \begin{equation}
    	I_{\text{bulk}} = - \int{ \mathbf{L}},
    	\label{eq_IE}
    \end{equation}
    where
    \begin{align}
        \mathbf{L}&=\mathbf{L}_{\mathrm{G}}+\lambda (\tau )\mathrm{d}\tau \land \left( \sqrt{\gamma}\left( \mathrm{d}^{D-1}x \right) _{\tau}-v\sqrt{\gamma}\left( \mathrm{d}^{D-1}x \right) _{\tau} \right),
    	\label{eq_Lag}
        \\
        v&=V\left( \int_{\partial \Sigma _{\tau}}{\sqrt{\gamma}\left( \mathrm{d}^{D-1}x \right) _{\tau}} \right) ^{-1}.
        \label{eq_nonlocalv}
    \end{align}
    Here, $\mathbf{L}_{\text{G}}$ represents the purely gravitational part, which we assume to be generally covariant. In the case of Einstein gravity, it is given by
    \begin{equation}
        \mathbf{L}_{\text{G}} =\frac{1}{16\pi}\sqrt{g}R\mathrm{d}^Dx.
    \end{equation}
    To express the action in the form of a $D$-form, the nonlocal term $v$ is introduced. However, the second term in Eq.~\eqref{eq_Lag} does not constitute a strictly proper $D$-form. This reflects an explicit breaking of the full diffeomorphism invariance of the bulk manifold $M$, reducing the symmetry effectively to independent diffeomorphisms acting on the Euclidean time circle $\mathbb{S}^1$ and on the spatial slice $\Sigma_\tau$. Consequently, directly applying the covariant phase-space formalism to the full Lagrangian density \eqref{eq_Lag} involves certain technical complications. Nonetheless, it should be emphasized that our primary objective is to establish the relationship between the Noether charge and the geometric contributions to the Euclidean action. Since all geometries under consideration satisfy the volume constraint, only the purely gravitational term $\mathbf{L}_{\mathrm{G}}$ gives a nonvanishing contribution to the Euclidean bulk action $I_E$. Therefore, it suffices to apply the covariant phase space formalism to the generally covariant gravitational part $\mathbf{L}_{\mathrm{G}}$ alone.

    By varying the Lagrangian density $\mathbf{L}_{\text{G}}$ with respect to $g_{ab}$, one finds
    \begin{equation}
    	\delta \mathbf{L_\text{G}}=\mathbf{E}\left[ g \right] ^{ab}\delta g_{ab} +\mathrm{d}\mathbf{\Theta }\left[ g,\delta g \right] ,
    	\label{eq_deltaL}
    \end{equation}
    where $\mathbf{E}[g]^{ab}$ denotes the Euler-Lagrange expression for the gravitational sector, and $\mathbf{\Theta}[g, \delta g]$ is the associated presymplectic potential. In the case of Einstein gravity, these are explicitly given by
    \begin{align}
    	&\mathbf{E}\left[ g \right] ^{ab}= E[g]^{ab}\sqrt{g}d^Dx= \frac{1}{16\pi} \left( \frac{1}{2}g^{ab}R-R^{ab} \right) \sqrt{g}\mathrm{d}^D x,
        \label{eq_EabForm}
        \\
    	&\mathbf{\Theta }\left[ g,\delta g \right] =\frac{1}{16\pi}\sqrt{g}\left( g^{ca}\nabla ^b-g^{ab}\nabla ^c \right) \delta g_{ab}\left( \mathrm{d}^{D-1}x \right) _c.
    \end{align}
    Next, consider the field transformation generated by a vector field $\xi^a$, under which the metric varies as
    \begin{equation}
    	\delta _{\xi}g_{ab}\equiv \mathcal{L} _{\xi}g_{ab}=\nabla _a\xi _b+\nabla _b\xi _a,
    \end{equation}
    corresponding to an infinitesimal diffeomorphism. Under this transformation, one finds,
    \begin{equation}
    	\delta _{\xi}\mathbf{L_{\text{G}}}= \mathcal{L} _{\xi} \mathbf{L_{\text{G}}}= \mathrm{d}\left( i_{\xi}\mathbf{L}_\text{G} \right).
    	\label{eq_deltaLSymtau}
    \end{equation}
    On the other hand, applying \eqref{eq_deltaL} gives
    \begin{equation}
       \delta_{\xi} \mathbf{L_\text{G}}=2 \mathbf{E}\left[ g \right] ^{ab} \nabla_a \xi_{b} +\mathrm{d}\mathbf{\Theta }\left[ g,\delta_\xi  g \right] ,
    \end{equation}
    Hence, we obtain
    \begin{equation}
        \mathrm{d}\left( i_{\xi}\mathbf{L}_\text{G} \right) = 2 \mathbf{E}\left[ g \right] ^{ab} \nabla_a \xi_{b} +\mathrm{d}\mathbf{\Theta }\left[ g,\delta_\xi  g \right].
        \label{eq_deltaL1eqdeltaL2}
    \end{equation}
    Taking the Euler-Lagrange derivative (see, e.g.,~\cite{Barnich:2018gdh,Ruzziconi:2019pzd} for related definitions) with respect to $\xi_a$ on both sides, the exact form terms drop out, leaving
    \begin{equation}
        \nabla _aE\left[ g \right] ^{ab}=0.
    \end{equation}
    This, in turn, implies that Eq.~\eqref{eq_deltaL1eqdeltaL2} can be rewritten as
    \begin{equation}
        \mathrm{d}\left( i_{\xi}\mathbf{L}_{\mathrm{G}}-\mathbf{M}\left[ \xi \right] -\mathbf{\Theta }\left[ g,\delta _{\xi}g \right] \right) =0,
    	\label{eq_deltaL2}
    \end{equation}
    where
    \begin{equation}
        \mathbf{M}\left[ \xi \right] =2\xi _a E\left[ g \right] ^{ab} \left( \mathrm{d}^{D-1}x \right) _b.
        \label{eq_Mexp}
    \end{equation}
   By the algebraic Poincar\'{e} lemma~\cite{Barnich:2018gdh,Ruzziconi:2019pzd}, locally there exists a Noether charge $\mathbf{Q}[\xi]$ such that
    \begin{equation}
        i_{\xi}\mathbf{L}_{\mathrm{G}}-\mathbf{M}\left[ \xi \right] -\mathbf{\Theta }\left[ g,\delta _{\xi}g \right] =-\mathrm{d}\mathbf{Q}\left[ \xi \right].
        \label{eq_RLdQ}
    \end{equation}
    It should be noted that the Noether charge $\mathbf{Q}\left[ \xi \right]$ is not uniquely defined; in what follows, we adopt the Noether-Wald charge~\cite{Iyer1994SomeProperties}. A direct computation shows that the Noether charge for Einstein gravity is
    \begin{equation}
    	\mathbf{Q}\left[\xi\right]\equiv -\frac{1}{8\pi}\sqrt{g}\nabla ^a\xi ^b\left( \mathrm{d}^dx \right) _{ab}.
    \end{equation}
    Substituting this into Eq.~\eqref{eq_RLdQ} with $\xi=\partial_\tau$ and integrating over $\Sigma_\tau$, one finds
    \begin{equation}
        -\int_{\Sigma _{\tau}}{i_{\partial _{\tau}}\mathbf{L}_\text{G}}=\int_{\partial \Sigma _{\tau}}{\mathbf{Q}\left[ \partial _{\tau} \right]}-\int_{\Sigma _{\tau}}{\mathbf{\Theta }\left[ g,\delta _{\partial _{\tau}}g \right]}-\int_{\Sigma _{\tau}}{\mathbf{M}\left[ \partial _{\tau} \right] }.
    \end{equation}
    Inspecting Eq.~\eqref{eq_Mexp} shows that the third term is tied to the $\tau\tau$- and $\tau i$-components of the gravitational field equations, namely, the Hamiltonian and momentum constraints. As Eq.~\eqref{eq_EOMforVC} makes clear, the volume constraint leaves these constraints unchanged. Hence, provided the Hamiltonian and momentum constraints are satisfied, even with the volume constraint imposed, the third term contributes nothing. If the Euclidean geometry further admits translational symmetry along the $\tau$-direction, the second term also vanishes. Under these considerations, the bulk gravitational action reduces to the boundary contribution $\mathbf{Q}[\partial_\tau]$ at $\partial M$. Performing the remaining integration over the $\tau$-direction, for Euclidean geometries that satisfy the volume constraint, we obtain
    \begin{equation}
    \begin{split}
        I_{\mathrm{bulk}}=-\int_M{\mathbf{L}}&=-\int_0^{\beta}{d\tau \int_{\Sigma _{\tau}}{i_{\partial _{\tau}}\mathbf{L}_\text{G}}}
        \\
        &=\int_0^{\beta}{d\tau \int_{\partial\Sigma _{\tau}}{\mathbf{Q}\left[ \partial _{\tau} \right]}}.
    \end{split}
    	\label{eq_IbulkQ}
    \end{equation}
    If $\partial\Sigma_\tau$ includes the bifurcation surface $\mathcal{B}$ (i.e., $\Sigma_\tau$ intersects the Killing horizon associated with $\partial_\tau$), then the boundary term in Eq.~\eqref{eq_IbulkQ} evaluated on $\mathcal{B}$ yields the Wald entropy $S_W$, which in general relativity reduces to one quarter of the horizon area.

    In the Euclidean action \eqref{eq_EuclideanAction}, the GHY term contributes only when the manifold $M$ has a boundary; if $M$ is boundaryless, the action is entirely determined by the bulk term. In this case, the Euclidean action reduces to a surface integral over $\partial\Sigma_\tau$, and when $\partial\Sigma_\tau = \mathcal{B}$, it is fixed solely by the horizon contribution $S_W$. For Einstein gravity, three-dimensional massive gravity, and Lovelock gravity, it has been shown~\cite{Jacobson:2022FVPartitionFunction,Tavlayan:2023FVHigherCurvatureTheory,Lu:2024FVLovelockTheory} that the on-shell Euclidean action coincides with the Wald entropy, giving the semiclassical partition function $\mathcal{Z}\sim e^{S_W}$.

    In the subsequent analysis of the Euclidean action, Eq.~\eqref{eq_IbulkQ} greatly simplifies the computation. In particular, as shown in Eqs.~\eqref{eq_SolforNC} or \eqref{eq_Nchid}, $N(r)$ is intrinsically complicated, making the evaluation of the bulk contribution for the metric \eqref{eq_ds2Nr} rather difficult. Equation~\eqref{eq_IbulkQ}, however, shows that it is sufficient to consider only the metric at $\partial\Sigma_\tau$.

    \subsection{VCEG contributions to the Euclidean action}\label{Sec_EuclideanFormalismIyerWaldEntropy}

\begin{figure}[h]
    \begin{center}
			\subfigure[$I_E$ vs. $\chi_c $\label{Fig_chicIe}]{\includegraphics[width=6 cm]{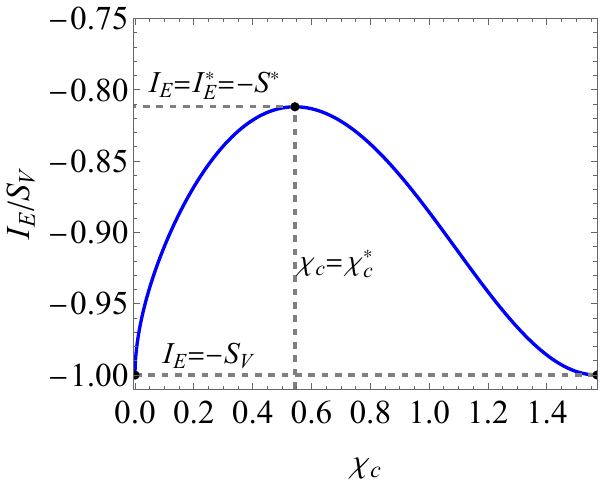}}
            \quad
			\subfigure[$I_E$ vs. $m $ \label{Fig_mIe}]{\includegraphics[width=6 cm]{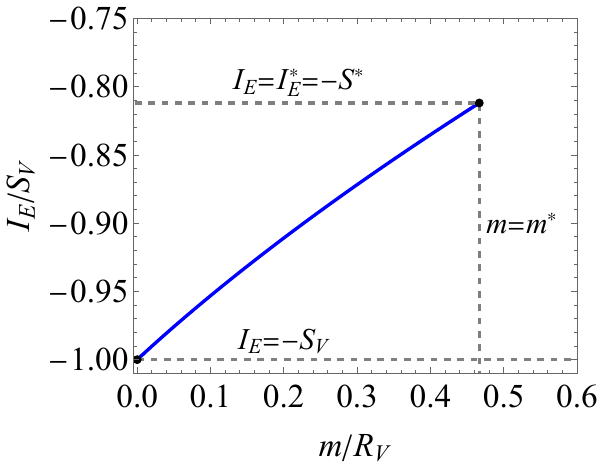}}
	\end{center}
        \caption{Euclidean action $I_E/S_V$ as a function of (a) $\chi_c$ and (b) $m/R_V$ in four dimensions. Here the volume constraint is $V = 4\pi R_V^{3}/3$, and $S_V \equiv \pi R_V^2$. The curves marked with an asterisk correspond to the critical configuration with $\chi_c = \chi_c^*$ and $m = m^*$.}
    \end{figure}

\begin{figure}[h]
    \begin{center}
			\subfigure[$I_E$ vs. $\chi_c $\label{Fig_chicIeD}]{\includegraphics[width=6 cm]{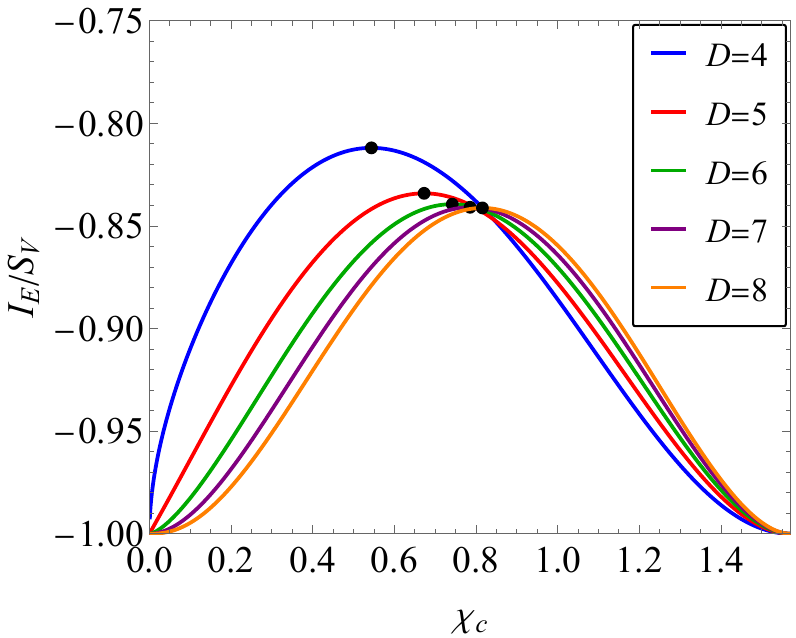}}
            \quad
			\subfigure[$I_E$ vs. $m $ \label{Fig_mIeD}]{\includegraphics[width=6 cm]{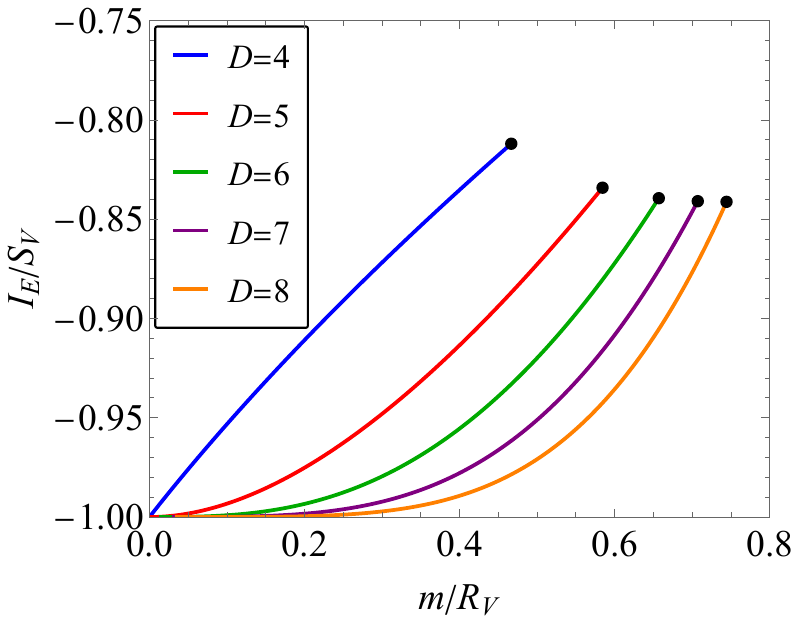}}
	\end{center}
        \caption{Euclidean action $I_E/S_V$ as a function of (a) $\chi_c$ and (b) $m/R_V$ in different dimensions. Here the volume constraint is $V = \Omega_d R_V^{d+1}/(d+1)$, and $S_V \equiv \Omega_d R_V^d/4$. Black dots indicate the critical configurations at $\chi_c = \chi_c^*$ and $m = m^*$.}
    \end{figure}

    We now turn to the contribution of the Euclidean geometry to the action. As discussed in Sec.~\ref{Sec_EuclideanFormalismIyerWaldEntropy}, the action receives contributions only from the horizon terms. In contrast to the previous case, however, here conical singularities are present, and one must add the corresponding contributions from $\chi=\chi_c$ and $\chi=\chi_e$. For Einstein gravity these take the form~\cite{Fursaev:1995ef,Solodukhin:1994yz,Solodukhin:2011gn,Morvan:2022ybp,Li:2022oup,Wu:2025jes}
   \begin{equation}
       -\frac{1}{4}\mathcal{A} _c\left( 1-\frac{\beta}{\beta _c} \right) -\frac{1}{4}\mathcal{A} _e\left( 1-\frac{\beta}{\beta _e} \right),
   \end{equation}
   where $\mathcal{A}_c$ and $\mathcal{A}_e$ denote the horizon areas at $\chi_c$ and $\chi_e$, respectively. The resulting Euclidean action is then
   \begin{equation}
   \begin{split}
       I_E&=-\frac{1}{4}\left( \mathcal{A} _c+\mathcal{A} _e \right)
       \\
       & = -\frac{\Omega _d}{4}\left( \frac{m^d}{(\cos \chi _c)^{2d/(d-1)}}+\frac{m^d}{(\cos \chi _e)^{2d/(d-1)}} \right).
   \end{split}
   \label{eq_Iechid}
    \end{equation}
    This expression depends solely on the horizon areas. Furthermore, the result is independent of $\lambda$, indicating that $I_E$ is a function only of $\chi_c$ (or, equivalently, of $m$). For the four-dimensional case, the corresponding curves are shown in Fig.~\ref{Fig_chicIe} and Fig.~\ref{Fig_mIe}, with the volume given by $V = 4\pi R_V^3 / 3$. For the $I_E-\chi_c$ curve, the point $\chi_c = \chi_c^*$ remains a stationary point (a local maximum), and the corresponding action equals $-S_V$, whereas $\chi_c = 0$ (or, equivalently, $\chi_c = \pi/2$) corresponds to another stationary point (a local minimum), for which the action equals $-S^*$. One observes that the Euclidean action is bounded from below, with the lower bound $I_E = -S_V \equiv -\pi R_V^2$ attained in the limit $\chi_c = 0$ or $\chi_c = \pi$ (equivalently $m = 0$). This indicates that turning on the mass parameter increases the Euclidean action, or equivalently, the free energy. In higher dimensions, a similar behavior is observed. The corresponding results are illustrated in Fig.~\ref{Fig_chicIeD} and Fig.~\ref{Fig_mIeD}.

    \section{Euclidean geometry as constrained instantons}\label{Sec_ConstrainedInstantons}

    As discussed earlier, the massive EVCEG possesses two horizons, with conical singularities arising at both $\chi = \chi_c$ and $\chi = \chi_e$. In general, these singularities cannot be simultaneously eliminated; they are removed only when $m$ attains its maximal value $m^*$, for which $\beta_c = \beta_e$. In contrast, the massless EVCEG features only a single horizon. For $ 0 < m < m^*$, the EVCEG inevitably exhibits conical singularities, indicating that the configuration is not in thermal equilibrium. This closely parallels the situation in pure dS spacetime and the SdS black hole. The Euclidean SdS geometry likewise possesses two horizons, each of which is generically associated with a conical singularity. These singularities can be simultaneously removed only in the Nariai limit (corresponding to the maximal black hole mass), where the two horizons coincide in the standard coordinates and their surface gravities become equal. However, near this degenerate limit, by introducing appropriate coordinates that resolve the degeneracy, the horizons can be effectively separated, allowing the conical singularities at both to be eliminated via suitable choices of the Euclidean time periodicity~\cite{Nariai:1951,Ginsparg:1982rs}. In contrast, the massless limit of the SdS black hole reduces to pure dS spacetime with a single cosmological horizon. Taken together, these features establish a close structural analogy between the EVCEG configuration and the Euclidean SdS black hole.

    For EVCEGs with mass in the range $0 < m < m^*$, i.e., the non-critical massive EVCEGs, the presence of conical singularities implies that the extended Euclidean solutions are not genuine solutions of the gravitational field equations. In this section, we will show that such configurations should instead be interpreted as configurations at fixed mass parameter $m$, namely as mass-constrained instantons, in close analogy with the treatment of the Schwarzschild--de Sitter black hole discussed in Refs.~\cite{Draper:2022xzl,Morvan:2022ybp}.

    \subsection{Constrained instantons in gravity}

    An essential element in our analysis is the method of constrained instantons~\cite{Frishman:1978xs,Affleck:1980mp}, originally developed in quantum field theory. Recently, this idea has been further explored in the context of gravitational theories~\cite{Cotler:2020lxj,Cotler:2021cqa}. Constrained instanton configurations are particularly significant because, although they are not saddle points for the Euclidean action in the conventional sense, they still contribute to nonperturbative processes in quantum gravity. In Einstein gravity with a fixed-volume constraint, the massive EVCEG provides a concrete realization of such a constrained instanton.

    We begin with the path integral formulation of quantum gravity, where the fundamental variable is the metric field $g_{ab}$. The partition function can be rewritten as
    \begin{equation}
        \begin{split}
         \mathcal{Z}&=\int{\left[ \mathcal{D} g \right] \exp \left( -I_E[g] \right)}
         \\
         &=\int{d\mu}\int{\left[ \mathcal{D} g \right] \delta \left( \mathcal{C} \left[ g \right] -\mu \right) \exp \left( -I_E[g] \right)}
         \\
         &= \int{d\alpha \int{d\mu \int{\left[ \mathcal{D} g \right] \exp \left( -I_E[g]+\alpha \left( \mathcal{C} \left[ g \right] -\mu \right) \right)}}}.
         \end{split}
    \end{equation}
    In the second line, we have inserted a constraint $\mathcal{C}[g]$ together with an auxiliary integration variable $\mu$; this step is harmless. In the third line, the Dirac delta function has been represented as an integral of $\alpha$ along the imaginary axis. The resulting effective action now depends on the gravitational field $g_{ab}$ as well as on the additional parameters $\alpha$ and $\mu$. The corresponding equations follow from varying the exponent with respect to $g_{ab}$ and $\alpha$,
    \begin{equation}
        \frac{\delta I_E\left[ g \right]}{\delta g_{ab}}+\alpha \frac{\delta \mathcal{C} \left[ g \right]}{\delta g_{ab}} =0,\quad \mathcal{C} \left[ g \right] =\mu.
        \label{eq_GCgalphaeq}
    \end{equation}
    These equations define the gravitational field equations under the constraint, where $\alpha$ acts as a Lagrange multiplier. If one further varies with respect to $\mu$, the condition $\alpha = 0$ arises, which restores the first equation to the ordinary, unconstrained Einstein equations. However, if we ignore the variation with respect to $\mu$ and solve only Eq.~\eqref{eq_GCgalphaeq}, we obtain a family of solutions $g_{ab}^{(\mu)}$ and corresponding multipliers $\alpha^{(\mu)}$. Whenever $\alpha^{(\mu)} \neq 0$, the resulting geometry is referred to as a constrained instanton. Within the saddle-point approximation, substituting these constrained instantons back into the path integral yields
    \begin{equation}
       \mathcal{Z} \simeq \int{d\mu \exp \left( -I_E[g^{\left( \mu \right)}] \right)}.
       \label{eq_ZVCmu}
    \end{equation}
    Thus, the partition function is expressed as a sum over geometries labeled by the constraint parameter $\mu$. In general, the extrema of $I_E [g^{(\mu)}]$ with respect to $\mu$ correspond to the ordinary (unconstrained) saddle points, which are characterized by $\alpha^{(\mu)} = 0$. Inspired by this, one can infer that the extrema along the curve of the Euclidean action $I_E[g^{(\mu)}]$ as a function of $\mu$ correspond to genuine saddle points, while the constrained instantons play the role of intermediate configurations connecting these saddles.

    It is worth emphasizing that the action \eqref{eq_EuclideanAction} can be regarded as the Einstein--Hilbert action supplemented by a constraint, where the Lagrange multiplier $\mu$ enforces a fixed spatial volume $V$. Within this setup, the massless and critical EVCEGs can be identified as constrained instantons of Einstein gravity. However, our primary focus is on the volume-constrained Einstein action itself, which defines the variational principle relevant for the path integral. In this framework, the massless and critical EVCEGs correspond to genuine saddle points, as they extremize the action subject only to the volume constraint. By contrast, non-critical massive EVCEGs do not arise as extrema of this action and therefore should not be regarded as instantons in this sense, unless an additional constraint associated with the mass is imposed. We now make this distinction explicit by constructing the non-critical massive EVCEGs as constrained instantons with fixed mass, implemented by fixing the extrinsic curvature on a timelike hypersurface, as described below.

    For the non-critical massive EVCEG, although the unavoidable conical singularities prevent it from being a true saddle point of the action \eqref{eq_EuclideanAction}, such a geometry should be interpreted as a constrained instanton with fixed mass. Specifically, following the discussion of the SdS black hole in Refs.~\cite{Draper:2022xzl,Morvan:2022ybp}, we divide the EVCEG into two manifolds by introducing a hypersurface at $\chi = \chi_0$, such that the two conical singularities at $\chi = \chi_c$ and $\chi = \chi_e$ are separated onto the two manifolds. On this hypersurface, we fix the extrinsic curvature $k$ as the constraint condition, which is equivalent to fixing the mass parameter. In this way, the conical singularities on the two manifolds can be removed by assigning appropriate periods to the Euclidean time coordinate on each manifold. Thus, the non-critical massive EVCEG is interpreted as a mass-constrained gravitational instanton. In the Hamiltonian formulation, the contributions from the bulk and the boundaries vanish under the imposed constraints. However, since the horizons are not boundaries where GHY terms are a priori included, one must compensate for this by restoring the GHY contributions at the horizons. As a result, the total contribution to the action is determined by the GHY terms evaluated at the horizons, which yield the corresponding area contributions. This result is also consistent with our expression \eqref{eq_Iechid}.

    From another viewpoint, the ensemble-averaged theory of black holes, it is also natural to include configurations with conical singularities in the path integral. As discussed in Refs.~\cite{Cheng:2024hxh,Cheng:2024efw,Ali:2024adt,Liu:2025iei}, for a given ensemble temperature, a Euclidean black hole typically possesses conical singularities unless the ensemble temperature exactly matches the Hawking temperature. In this case, the Euclidean action becomes a function of the horizon radius $r_h$, and the stationary points of this action with respect to $r_h$ correspond precisely to regular (singularity-free) geometries. Moreover, when the ensemble average is taken, the Hawking-Page transition for the SAdS black hole, as well as the small-large black hole transition for the Reissner-Nordström AdS black hole, naturally emerge as semiclassical approximations in the regime where the system size is much larger than the Planck length. Beyond the semiclassical limit, the system becomes a superposition of different geometries, and the averaged quantities deviate from standard black hole thermodynamics. In our case, for the non-critical massive EVCEG, there generally exist unavoidable conical singularities under a fixed ensemble temperature, located respectively on the two horizons. From the ensemble-averaged perspective, although such configurations possess conical defects, they should still contribute to the gravitational path integral. Furthermore, as discussed above, the dominant contribution comes from the $m = 0$ configuration, which corresponds to a stable point.

    Therefore, whether viewed from the perspective of constrained instantons or from the ensemble-averaged theory, the inclusion of these constrained instanton solutions in the semiclassical path integral is well justified. In analogy with the SdS case, we may identify the constraint parameter $\mu$ with the ratio $M/M^*$, whose range is $[0,1]$. This choice is motivated by the SdS case and is intended to facilitate a direct comparison with the analysis in Ref.~\cite{Morvan:2022ybp}. Still, a supplementary remark is in order. In the present treatment, we have assumed that, when $\mu$ is treated as the integration variable, the integrand is dominated solely by the exponential term, while the prefactor is treated as constant, as in Eq.~\eqref{eq_ZVCmu}. In general, this prefactor receives contributions from fluctuations and zero modes. Even so, we expect that as a conceptual and approximate description of the instanton contribution, the expression in Eq.~\eqref{eq_ZVCmu} provides a reasonable approximation.

\subsection{Ensemble interpretation and quantum decay rate}

    As discussed before, both the gravitational instantons and the true saddle points are included in the partition function. The saddle points correspond to the massless EVCEG and the critical EVCEG, representing the Euclidean minimum and maximum of the action, respectively. The non-critical massive EVCEGs, interpreted as gravitationally constrained instantons, can then be regarded as an intermediate configuration connecting these two extrema. Therefore, in the absence of quantum effects, the massless EVCEG is classically stable. However, once quantum fluctuations are taken into account, a decay process may occur.

\begin{figure}[h]
        \centering
        \includegraphics[width=6cm]{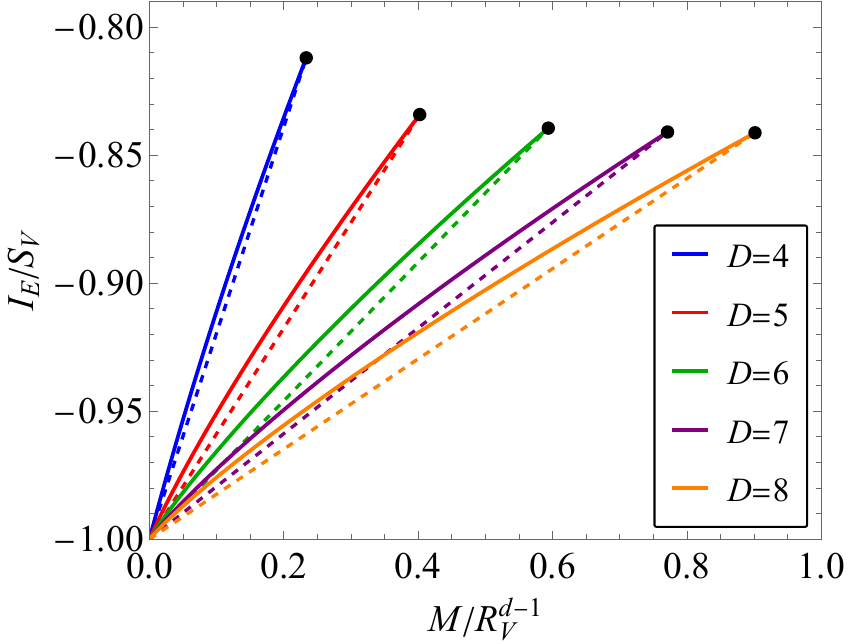}
        \caption{Relations of the Euclidean action with $M$ in different spacetime dimension. Black dots indicate the critical configurations at $\chi_c = \chi_c^*$, $m = m^*$ and $M =M^*$.}
       \label{Fig_md1IE}
    \end{figure}

    First, let us consider the contribution of the EVCEG to the partition function. Since the integration is over the constraint parameter $\mu = M/M^*$ and the Euclidean action \eqref{eq_Iechid} depends on $\mu$ in a rather complicated manner, the resulting integral cannot be evaluated in analytic expression. To further address the difficulty of evaluating the integral, it is reasonable to adopt an approximate approach. As a first step, let us analyze the behavior of the Euclidean action $I_E$ in the regime of small mass constant $M$. To clarify this behavior, let us take $\chi _c=\frac{\pi}{2}-\epsilon ^{d-1}$, where $\epsilon$ is a small dimensionless parameter. Using the condition $N(\chi_e) = 0$, we find $\chi _e=-\frac{4}{d-1}\epsilon ^4+o(\epsilon^4)$. With the aid of the volume constraint \eqref{eq_Vcd} and the Euclidean action \eqref{eq_Iechid}, we then obtain
    \begin{equation}
    \begin{split}
        \frac{m}{R_V}&=\epsilon ^2-\frac{3d+1}{12\left( d-1 \right)}\epsilon ^{2d}+  o\left(\epsilon^{2d}\right), 
        \\
        \frac{I_E}{S_V}&=1-\frac{1}{4}d\epsilon ^{2d-2}+o\left(\epsilon^{2d-2}\right),
    \end{split}
    \end{equation}
    where $S_V$ is defined as one quarter of the area of a $d$-sphere with radius $R_V$, i.e. $S_V = \Omega_d R_V^d / 4$. Noting that $m^{d-1}= 16\pi M/(d \Omega_d)$, eliminating $\epsilon$ gives
    \begin{equation}
    \begin{split}
        \frac{I_E}{S_V}&=-1+\frac{1}{4}d\left( \frac{m}{R_V} \right) ^{d-1}+o\left( \left( \frac{m}{R_V} \right) ^{d-1} \right) 
        \\
        &=-1+\frac{4\pi}{\Omega _d}\frac{M}{R_{V}^{d-1}}+o\left( \frac{M}{R_{V}^{d-1}} \right),
    \end{split}        
        \label{eq_IesmallM}
    \end{equation}
    which suggests that, at leading order, a linear relation arises between $I_E$ and $M$. Accordingly, we adopt the following approximate relation
    \begin{equation}
    \begin{split}
        I_E &\approx -S_V+\frac{S_V-S^*}{M^*}M=-S_V+\Delta S\mu ,
        \\
        \Delta S &\equiv S_V - S^*,
    \end{split}
    \label{eq_IEapprox}
    \end{equation}
    where the entropy $S^*$ represents that of the critical EVCEG, and the entropy difference $\Delta S$ is positive. The corresponding linear approximation (dashed lines) and the exact results (solid lines) are shown in Fig.~\ref{Fig_md1IE}.
    Under this approximation, the corresponding partition function can be evaluated as
    \begin{equation}
    \begin{split}
        \mathcal{Z} \simeq \int{d\mu \exp \left( -I_E[g^{\left( \mu \right)}] \right)} &\approx \int_0^1{d\mu \,\,e^{S_V-\Delta S\mu}}
       \\
       &=\frac{e^{S_V}}{\Delta S}\left( 1-e^{-\Delta S} \right).
    \end{split}
    \label{eq_Zapprox}
    \end{equation}
    It is evident that the partition function depends only on the entropies associated with the two limiting configurations, namely massless EVCEG and critical EVCEG. We note that for $m = m^*$, the corresponding $\chi_c = \chi_c^*$ and $\chi_e = \chi_e^*$ are determined as the roots of $H(\chi) = 0$, and therefore do not depend on the volume radius $R_V$. From the volume constraint~\eqref{eq_Vcd}, it follows that $m^* \propto R_V$, with a proportionality factor that depends on the dimension of geometry. Restoring the Planck length in the expression for the entropy, one finds $S^* \propto (R_V/\ell_p)^d$ and $\Delta S \propto (R_V/\ell_p)^d$, similar to $S_V$, though with different numerical coefficients. Consequently, the partition function~\eqref{eq_Zapprox} depends only on the dimensionless ratio $R_V/\ell_p$ and on the dimension $D$.

    When the ratio $R_V/\ell_p$ becomes either much smaller or much larger than unity, the partition function can be approximated as
    \begin{equation}
        \mathcal{Z} \approx
            \begin{cases}
            e^{S_V}\left( 1-\frac{1}{2} \Delta S \right) ,
            &	\text{for}\; R_V/\ell _p\ll 1 \;,
            \\
            e^{S_V} \frac{ 1}{\Delta S} ,
            &	\text{for}\; R_V/\ell _p\gg 1.
            \end{cases}
    \end{equation}
    It is evident that when the radius $R_V$ is much larger than the Planck length $\ell_p$, the partition function acquires only a multiplicative correction factor. In contrast, when $R_V \ll \ell_p$, gravitational effects become significant and $S_V$ itself becomes small. In this regime, one can further expand
    \begin{equation}
    \begin{split}
        e^{S_V}\left( 1-\frac{1}{2}\Delta S+\mathcal{O} (\Delta S^2) \right) &\approx e^{S_V-\frac{1}{2}\Delta S+\mathcal{O} (\Delta S^2)}
        \\
        &=e^{\frac{1}{2}\left( S_V+S^* \right) +\mathcal{O} (\Delta S^2)},
    \end{split}
    \end{equation}
    indicating that the effective entropy is corrected to the mean value of the two entropies corresponding to massless EVCEG and critical EVCEG. This indirectly indicates that the saddle-point approximation ceases to be reliable in this regime, as nearly all configurations contribute with comparable weight, a manifestation of strong quantum fluctuations. Returning to the unapproximated expression of the partition function, Eq.~\eqref{eq_Zapprox}, we note that the entropies of EVCEGs are smaller than $S_V$ and larger than $S^*$, resulting in an effective entropy $\log \mathcal{Z}$ that is smaller than $S_V$. Consequently, as $R_V/\ell_p$ decreases (i.e., as quantum fluctuations become increasingly significant), the entropy of the gravitational system drops below $S_V$ and approaches the lower bound $(S_V + S^*)/2$. Equivalently, this entropy reduction can be interpreted as an increase in the effective gravitational coupling $G_{\text{eff}}$ under a fixed area law. Taking $\ell_p$ as the reference energy scale, this implies that at higher energies (smaller $R_V$), the effective coupling $G_{\text{eff}}$ becomes larger.

    Below, we briefly consider the decay probability from the massless EVCEG to massive EVCEGs, which can be approximately expressed as
    \begin{equation}
    \Gamma \approx M^*\int_0^1{d\mu \,\,e^{-\Delta S\mu ^{d-1}}} =\frac{M^*}{\Delta S}\left( 1-e^{-\Delta S} \right).
    \end{equation}
    In the weak-gravity regime, where $R_V/\ell_p \gg 1$, one finds
    \begin{equation}
    \Gamma \approx M^*\frac{1}{\Delta S} = \frac{M^*}{S_V-S^*}=\#\frac{1}{R_V},
    \label{eq_decayrate}
    \end{equation}
    where "$\#$" denotes a dimensionless coefficient independent of $R_V$, which can be explicitly determined from the expressions of $S_V$, $M^*$ and $S^*$. The numerical values of $\#$ for several spacetime dimensions are presented in Table~\ref{Table_dacaycoefficient}. Since the temperature scales as $T \propto 1/R_V$ (with $\lambda m^2$ treated as dimensionless), the decay rate in this limit is therefore proportional to the temperature.

    \begin{table}[htbp]
        \caption{Numerical values of the coefficient $\#$ in the decay probability for different dimensions $D$. Here $\#$ is given in Eq.~\eqref{eq_decayrate}.}
        \label{Table_dacaycoefficient}
        \begin{ruledtabular}
        \begin{tabular}{ccccccc}
        $D$ & $4$ & $5$ & $6$ & $7$ & $8$ &...\\
        \hline
        $\#$ & $0.395$ & $0.491$ & $0.562$ & $0.625$ & $0.686$ &...\\
        \end{tabular}
        \end{ruledtabular}
    \end{table}

    In comparison with the analysis of SdS spacetime in Ref.~\cite{Morvan:2022ybp}, the EVCEG exhibits strikingly similar features. As shown in Eq.~\eqref{eq_IesmallM}, the Euclidean action exhibits a similar functional dependence on the mass constant $M$ at leading order, and in the weak-gravity limit the decay rate is likewise proportional to the temperature. While the SdS spacetime reflects the influence of a cosmological constant, the EVCEG corresponds to a gravitational configuration constrained by a fixed volume. This analogy indicates that the volume constraint plays a role similar to that of the cosmological constant, leading to comparable physical properties.


	\section{Conclusion and discussion}~\label{Sec_Discussion}

	We have examined the Euclidean action and partition function of gravity under a fixed-volume constraint. This work extends the analysis of Ref.~\cite{Jacobson:2022FVPartitionFunction}, in which the dominant saddle corresponds to a spherically symmetric configuration with a vanishing mass function. We have constructed constrained gravitational instantons with nonzero mass parameters (massive EVCEGs) and evaluated their contributions to the Euclidean action. By incorporating these massive EVCEGs, we clarify how the volume constraint shapes the space of admissible gravitational configurations and determines their relative weights in the ensemble.

    Within the covariant phase space framework, as discussed in the Supplemental Material of Ref.~\cite{Jacobson:2022FVPartitionFunction}, we review that even under a fixed volume constraint, configurations satisfying the Hamiltonian and momentum constraints contribute to the action solely through horizon and boundary terms: the horizon accounts for the Wald entropy, while the boundary contributes to the energy. Consequently, for geometries without boundaries, the action in general relativity is universally determined by the horizon area, with the total energy vanishing. This observation reinforces the interpretation that the partition function encodes the dimension of the Hilbert space, with the relevant degrees of freedom effectively localized on the horizon. This picture is reminiscent of the holographic principle. In the case of the massless VCEG, the Euclidean action reduces entirely to one quarter of the horizon area, a result similar to that of pure dS space.

    For Einstein gravity with a volume constraint, we obtain on-shell solutions corresponding to VCEGs with nonvanishing mass parameter. These geometries contain both a boundary at $r = m$ and a horizon. When the mass parameter is nonzero, the VCEG can be further analytically extended, leading to the EVCEG, in which the geometry no longer possesses a boundary but instead contains two horizons. Conversely, the VCEG can be regarded as the portion of the EVCEG with $\chi \ge 0$, corresponding to a manifold with boundary. Each of these horizons generically exhibits a conical singularity, and it is not possible to choose a single period of the Euclidean time that simultaneously removes both singularities, except in the critical case $m = m^*$ (i.e., the critical EVCEG). Therefore, for $0 < m < m^*$, the extended configurations (non-critical EVCEGs) do not correspond to genuine on-shell geometries. For the EVCEG, the Euclidean action is given by one quarter of the sum of the areas of the two horizons, in agreement with the result for the Schwarzschild--de Sitter black hole~\cite{Morvan:2022ybp}. Moreover, the following properties are also analogous to those of the SdS black hole: as shown in Eq.~\eqref{eq_IesmallM}, when the mass constant $M$ is treated as a small parameter, the subleading term of the Euclidean action is linear in $M$; and in the classical limit the corresponding decay rate is proportional to the temperature $T$ (as shown in Eq.~\eqref{eq_decayrate}).

    From a topological and geometrical perspective, the massless EVCEG and the Euclidean dS static patch are analogous, both being topologically equivalent to $\mathbb{S}^{d+2}$. In the case of nonzero mass, the massive EVCEG and the Euclidean SdS black hole each contain two horizons, which may exhibit conical singularities, and in general it is not possible to simultaneously remove both singularities. Only when the mass parameter of the massive EVCEG reaches its maximal value, corresponding to the maximal mass of the Euclidean SdS black hole, do the geometries reduce to the critical EVCEG and the Nariai geometry, respectively. Moreover, the metric of the massive EVCEG can be expressed as a warped product of $\mathbb{S}^2$ and $\mathbb{S}^d$. Since the radii of both factors remain nonvanishing throughout the geometry, the underlying manifold is topologically $\mathbb{S}^2 \times \mathbb{S}^d$, in agreement with the topology of the Euclidean SdS static patch.

   Before proceeding further, it is important to note that the above discussion applies to dimensions $D \ge 4$. The case of $D = 3$ is special, and the corresponding results are presented in Appendix~\ref{Appendix_threedimensioncase}. As shown there, in three dimensions neither the massless nor the massive VCEG requires any further extension, and the geometry is defined over the radial interval $r \in [0, r_c]$. For any value of the mass parameter $m$, VCEG contains only a single horizon, and its topology is that of $\mathbb{S}^3$. These features closely parallel those of the three-dimensional Euclidean SdS black hole. Moreover, just as Euclidean black hole, the contribution of the VCEG to the Euclidean action is determined solely by the area term. Thus, in $D = 3$, the two systems display a clear similarity.

    Both topological and Euclidean action analyses reveal that the EVCEG closely parallels the SdS black hole. These results suggest that the volume constraint effectively plays a role analogous to that of a cosmological constant. It is instructive to consider the Euclidean actions of the two cases. More specifically, the volume constraint and the cosmological constant terms can be written as
    \begin{equation}
        -\lambda(\tau) \sqrt{\gamma},\quad \frac{1}{8\pi}\Lambda \sqrt{g} = \frac{1}{8\pi}\Lambda N \sqrt{\gamma},
        \label{eq_comparelL}
    \end{equation}
    where we have omitted the volume term $V$, as it does not affect the equations of motion for the gravitational field. First, the EVCEG arises for $\lambda < 0$, while the dS geometry corresponds to $\Lambda > 0$. Therefore, the two share the same overall sign convention, a similarity that ensures the resulting geometry is as compact as possible. Second, the difference in the lapse function between the two cases implies that the corresponding gravitational solutions are not identical. Typically, for stationary solutions, the lapse function depends on the coordinate $r$. However, this difference does not play a crucial role in the present discussion, particularly in the analysis of the topology and the Euclidean action. Nevertheless, achieving a complete equivalence between the two terms in Eq.~\eqref{eq_comparelL} would require promoting the cosmological constant to a spacetime-dependent field $\Lambda(x)$, whose effect would precisely cancel that of the lapse function. Such a result may be realized from the viewpoint of symmetry breaking, where introducing an additional effective potential for $\Lambda(x)$ allows it to acquire the desired functional form.

    Upon reviewing the Euclidean action \eqref{eq_EuclideanAction}, it is clear that it consists solely of contributions from Einstein gravity and the volume constraint. The corresponding constrained instanton exhibit geometric features reminiscent of the SdS black hole and wormhole configurations. This observation suggests that certain geometric features typically associated with de Sitter spacetime, in particular the presence of a cosmological horizon, may in principle arise from an effective volume constraint rather than from a fundamental cosmological constant. This perspective may provide an alternative viewpoint on the origin of horizon structures in gravitational systems.

	\acknowledgments
    We are grateful to Xiang-Cheng Meng for valuable help and useful discussions. This work was supported by the National Natural Science Foundation of China (Grants No. 12475055, No. 12405073, and No. 12247101), the Fundamental Research Funds for the Central Universities (Grant No. lzujbky-2025-jdzx07), the Natural Science Foundation of Gansu Province (No. 22JR5RA389, No. 25JRRA799), and Natural Science Foundation of Tianjin (No. 25JCQNJC01920).

    \appendix

    \section{Special Case: three dimensions}\label{Appendix_threedimensioncase}

\begin{figure}[htp]
        \centering
        \includegraphics[width=6cm]{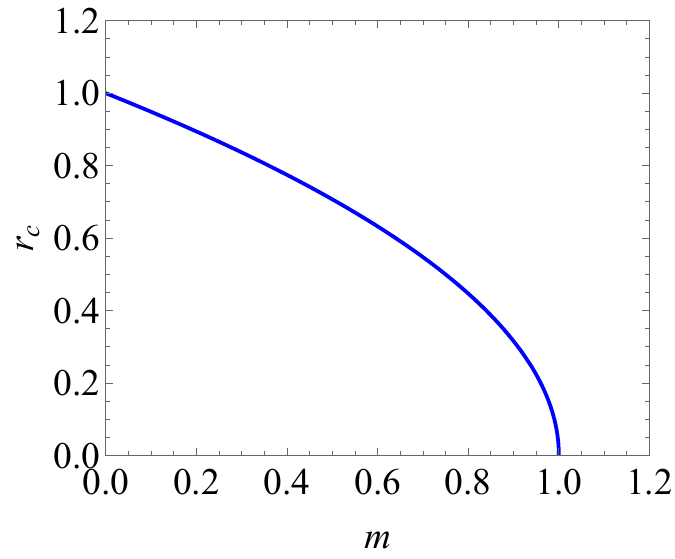}
        \caption{Relation between the mass parameter $m$ and $r_c$ at fixed volume. The volume is chosen to be that of a unit ball in two dimensions, $V = \pi$.}
       \label{Fig_3dMrc}
    \end{figure}

\begin{figure}[htp]
        \centering
       \includegraphics[width=6cm]{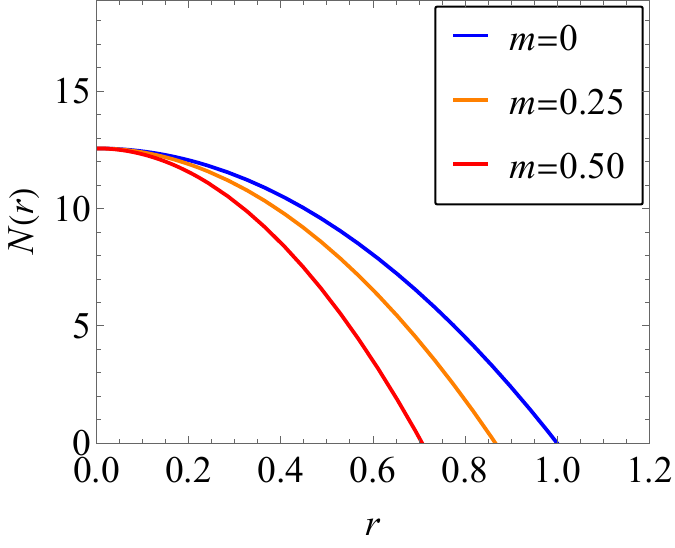}
        \caption{Functional profiles of $N(r)$ at fixed volume $V = \pi$. The blue, orange, and red curves correspond to $m = 0$ ($r_c = 1$), $m = 0.25$ ($r_c \approx 0.93$), and $m = 0.50$ ($r_c \approx 0.84$), respectively. Throughout the plot, the parameter $\lambda$ is fixed at $-1$.}
       \label{Fig_3drN}
    \end{figure}

    Although many of the conclusions remain similar in higher dimensions, for instance the appearance of cutoff regions in the spacetime due to mass excitations, the situation in three dimensions is markedly different. In analogy with Sec.~\ref{Sec_MassExcitation}, one may also discuss the TOV equation in the three-dimensional case. Here, however, we focus on the configuration in which the energy density $\rho(r)$ vanishes, as this corresponds to the solution of the field equations obtained by varying the action~\eqref{eq_EuclideanAction}. Concretely, the components of the energy-momentum tensor are given by $T_{ab} = -\rho(r)n_a n_b + p(r)\gamma_{ab}$, where
    \begin{equation}
        \rho \left( r \right) =0,\quad p\left( r \right) =\frac{m-1}{4\pi \left( r_{c}^{2}-r^2 \right)}.
    \end{equation}
    The corresponding gravitational solution takes the form,
    \begin{equation}
        N\left( r \right) =\frac{4\pi \lambda}{m-1}\left( r_{c}^{2}-r^2 \right) ,\quad M=\frac{m}{8},
    \end{equation}
    leading to the metric,
    \begin{equation}
        ds^2=\frac{16\pi ^2\lambda ^2}{\left( 1-m \right) ^2}\left( r_{c}^{2}-r^2 \right) ^2d\tau ^2+\frac{1}{1-m}dr^2+r^2d\phi ^2.
        \label{eq_3dmetric}
    \end{equation}
    The radial metric component $g_{rr}$ is independent of $r$, but it is required to be positive, which implies that the mass parameter $m$ must be smaller than 1. In analyzing the solution, we have assumed in advance that $N(r)>0$, which in turn requires $\lambda<0$, consistent with the case of $D \geq 4$. However, unlike in higher dimensions, the radial coordinate here ranges over $[0,r_c]$. It is important to note that as $r\to 0$, if one naively sets the periodicity of $\phi$ to $2\pi$, a conical singularity would arise at $r = 0$. To avoid this singularity, the periodicity of $\phi$ must instead be chosen as $2\pi /\sqrt{1-m}$. At this stage, the spatial volume constraint can be expressed as
    \begin{equation}
        V=\frac{1}{1-m}\pi r_{c}^{2}.
    \end{equation}
    Accordingly, the relation between $m$ and $r_c$ dictated by the volume constraint is shown in Fig.~\ref{Fig_3dMrc}. In contrast to the case of $d \geq 4$, here $r_c$ is a monotonically decreasing function of $m$, and the relation between $r_c$ and $m$ does not asymptotically approach the identity line at large $m$. Furthermore, under the volume constraint, the functional profiles of $N(r)$ for different values of $m$ are displayed in Fig.\ref{Fig_3drN}. These plots illustrate that varying $m$ effectively corresponds to changing the cutoff radius $r_c$.

    Considering the contribution of the metric \eqref{eq_3dmetric} to the action, we readily obtain
    \begin{equation}
        I_E=-\frac{1}{2\sqrt{1-m}}\pi r_{c} = -\frac{1}{2}\pi R_V,
    \end{equation}
    where $R_V$ is defined through the volume $V = \pi R_V^2$. It can be seen that, in three dimensions, a nonzero value of $m$ is in fact trivial. Indeed, by rescaling the coordinates as $r \rightarrow r\sqrt{1 - m}$ and $\phi \rightarrow \phi / \sqrt{1 - m}$, the metric~\eqref{eq_3dmetric} is brought back to the case with $m = 0$. In the above discussion, the conical singularity at $r=0$ is avoided by appropriately adjusting the periodicity of $\phi$. If instead one insists on fixing the periodicity of $\phi$ to $2\pi$ in the metric~\eqref{eq_3dmetric}, a conical singularity will arise at $r=0$. In this case, both the curvature and the energy density acquire delta-function contributions,
    \begin{equation}
            R \sim {T^0}_0 \sim \left(1 - \sqrt{1 - m}\right)\delta(r).
     \end{equation}
     This corresponds to the well-known conical singularity solution in three-dimensional gravity constructed by ’t Hooft, Deser, and Jackiw~\cite{Deser:1983tn,Carlip:1998uc}. Physically, this configuration does not represent a vacuum solution under the volume constraint, but rather describes the presence of a point particle located at $r=0$. Consequently, the conical singularity cannot be removed by a coordinate transformation.
    
	

\end{document}